\documentclass[aps,showpacs,preprint,preprintnumber,nofootinbib,amsmath,amssymb,ascmac,12pt]{revtex4}
\usepackage{bm}
\usepackage[dvipdfm]{graphicx}
\begin{document}
\title{\vspace*{1cm}
Stability analysis of Squashed Kaluza-Klein Black Holes\\ with Charge
}
\author{
       ${}^{1}$Ryusuke Nishikawa \footnote{E-mail:ryusuke@sci.osaka-cu.ac.jp}
and
       ${}^{1}$Masashi Kimura \footnote{E-mail:mkimura@sci.osaka-cu.ac.jp}
}
\affiliation{
${}^{1}$Department of Mathematics and Physics,
Graduate School of Science, Osaka City University,
3-3-138 Sugimoto, Sumiyoshi, Osaka 558-8585, Japan
\vspace{2cm}
}
\begin{abstract}
We study gravitational and electromagnetic perturbation around 
the squashed Kaluza-Klein black holes with charge. 
Since the black hole spacetime focused on this paper
have $SU(2) \times U(1) \simeq U(2)$ symmetry,
we can separate the variables of the equations for perturbations by using
Wigner function $D^{J}_{KM}$ which is the irreducible representation of the symmetry.
In this paper, we mainly treat $J=0$ modes which
preserve $SU(2)$ symmetry.
We derive the master equations for the $J=0$ modes and discuss the stability of these modes.
We show that the modes of $J = 0$ and $ K=0,\pm 2$ and the modes of $K = \pm (J + 2)$
are stable against small perturbations
from the positivity of the effective potential.
As for $J = 0, K=\pm 1$ modes,
since there are domains where the effective potential
is negative except for maximally charged case,
it is hard to show the stability of these modes in general.
To show stability for $J = 0, K=\pm 1$ modes in general is open issue.
However, we can show the stability for $J = 0, K=\pm 1$ modes in maximally charged case
where the effective potential are positive out side of the horizon.
\end{abstract}

\preprint{OCU-PHYS 328}
\preprint{AP-GR 75}
\pacs{04.50.+h, 04.70.Bw}

\date{\today}
\maketitle

\section{introduction}
In recent years, the studies on higher dimensional black holes have attracted much attention
in the context of the brane world scenario.
One of the main reason is that mini black hole creation in a particle collider was
suggested~\cite{Banks:1999gd,Dimopoulos:2001hw,Giddings:2001bu,Ida:2002ez,Ida:2005ax,Ida:2006tf}
based on some brane world model.
Such physical phenomena are expected to give us a piece of evidence for the existence of extra
dimensions and to draw some information towards quantum gravity.

It is known that higher-dimensional black objects admit various topology of horizon
unlike four-dimensional case.
In fact, there are many exotic solution of Einstein equation in higher dimensions
in addition to black holes, e.g. black rings, strings, Starns and
di-rings~\cite{Myers:1986un,Emparan:2001wn,Mishima:2005id,Figueras:2005zp,Pomeransky:2006bd,
Elvang:2007rd,Iguchi:2007is,Izumi:2007qx,Elvang:2007hs,Evslin:2008py}.
To predict what type of black objects are created in a particle collider, one of the important criterion
is their stability against small gravitational perturbations.
In~\cite{Ishibashi:2003ap}, Kodama and Ishibashi showed that
the equation of motion for perturbation of the maximally symmetric higher-dimensional black holes
reduces to single master equation for each mode.
They also showed the stability of the higher-dimensional Schwarzschild black holes.
Recently, much effort has been devoted to reveal the stability of 
higher-dimensional black holes with more complicated structure by many 
authors~\cite{Kodama:2003kk,Kunduri:2006qa,Murata:2008yx,Dias:2010eu,Kodama:2009rq,
Gregory:1993vy,Emparan:2003sy,Murata:2007gv,Astefanesei:2010bm}.

From a realistic point of view, the extra dimensions need to be 
compactified to reconcile the higher-dimensional gravity theory with 
our apparently four-dimensional world.
We call the higher-dimensional black holes on the spacetime with compact extra dimensions
Kaluza-Klein black holes.
In general, it is difficult to construct exact solutions describing Kaluza-Klein black holes
because of the less symmetry than the asymptotically flat case.
However, if we consider the spacetime with twisted extra-dimensions, we can 
construct such exact solutions, i.e. squashed Kaluza-Klein (SqKK) black holes~\cite{Dobiasch:1981vh,Gibbons:1985ac} 
in the class of cohomogeneity-one symmetry.
The topology of the horizon of this SqKK black
holes is $S^3$, while it looks like four-dimensional black holes with
a circle as an internal space in the asymptotic region.

Recently, much effort has been devoted
to reveal the properties of squashed Kaluza-Klein black holes.
In~\cite{Rasheed:1995zv,Larsen:1999pp,Ishihara:2005dp,Ishihara:2006iv,Wang:2006nw,
Yazadjiev:2006iv,Brihaye:2006ws,Ida:2007vi,Ishihara:2006ig,Yoo:2007mq,Kimura:2009gy,
Matsuno:2007ts,Nakagawa:2008rm,Tomizawa:2008hw,Tomizawa:2008rh,Stelea:2008tt,Tomizawa:2008qr,Gal'tsov:2008sh,
Allahverdizadeh:2009ay} generalizations of SqKK black holes are studied.
Several aspects of SqKK black holes are also discussed, e.g.
thermodynamics~\cite{Cai:2006td,Kurita:2007hu,Kurita:2008mj},
Hawking radiation~\cite{Ishihara:2007ni,Chen:2007pu,Wei:2009kg},
gravitational collapse~\cite{Bizon:2006ue},
behavior of Klein-Gordon equation~\cite{Radu:2007te},
stability for neutral case~\cite{Kimura:2007cr}\footnote{
Note that the stability of $J=0, K=\pm 1$ modes were not shown analytically
since there are some mistakes in the discussion in $J=0, K = \pm1$ modes in~\cite{Kimura:2007cr}.
It is hard to find S-deformation function to show stability for these modes analytically.
To find S-deformation function analytically is open issue.
However, the stability for these modes have been confirmed by numerical method in~\cite{Ishihara:2008re}.},
quasinormal modes~\cite{Ishihara:2008re,He:2008im},
geodetic precession~\cite{Matsuno:2009nz},
Kerr/CFT correspondence~\cite{Peng:2009wx}
and gravitational lensing effects~\cite{Liu:2010wh}.

In this paper, 
we study gravitational and electromagnetic perturbation around 
the squashed Kaluza-Klein black holes with charge~\cite{Ishihara:2005dp}
by extending the analysis in~\cite{Kimura:2007cr}.
SqKK black hole with charge
have the $SU(2) \times U(1)$ symmetry as in the neutral case,
so we can use the same technics used in~\cite{Hu:1974hh,Kimura:2007cr,Murata:2007gv} to analyze the perturbations of 
SqKK black holes with charge.
We expand the perturbation variables in terms of
Wigner function $D^{J}_{KM}$, which is the irreducible representation of
the symmetry characterized by three indices $J,M,K$.
Since the instability empirically appears in the lower modes, we mainly focus on
the perturbations in $J=0$ modes which preserve $SU(2)$ symmetry.
In this paper, we derive the master equations for $(J=0,M=0,K=0,\pm 1,\pm 2)$ modes which have $SU(2)$ symmetry
and the highest modes $(K=\pm (J+2) )$.
Using the effective potential functions in the master equations,
we discuss the stability for these modes.

There are several motivations to study the perturbation for SqKK black holes with charge.
One is to understand the property of the spacetime deeper
by using small perturbation as a probe such as quasinormal modes.
Next, the stability of SqKK black holes is
needed for the arguments in the physics around 
the black holes~\cite{Cai:2006td,Kurita:2007hu,Kurita:2008mj,Ishihara:2007ni,Chen:2007pu,Wei:2009kg,Ishihara:2008re,He:2008im,Matsuno:2009nz,Liu:2010wh}
to be meaningful.
Finally, it is also useful to consider the perturbation for SqKK black holes with charge to understand 
the effective theories reduced from unified theories based on string theory which
usually contains not only gravity but also gauge fields.

The organization of this paper is as follows.
In section~\ref{sec:SQBH}, 
we review the geometry of SqKK black holes with charge
and the formalism to classify metric perturbations based on the symmetry.
In section~\ref{sec:stability}, we derive the master equations for
master variables. By analyzing these equations, we discuss the stability of SqKK black holes with charge.
The final section is devoted to the discussion.

\section{Squashed Kaluza-Klein black holes with charge}\label{sec:SQBH}
In this section,
we review the geometry of SqKK black holes with charge~\cite{Ishihara:2005dp}
and the formalism to classify metric perturbations based on 
the symmetry~\cite{Hu:1974hh,Murata:2007gv,Kimura:2007cr}.
\subsection{Geometry of squashed Kaluza-Klein black holes with charge}\label{sec:SqKKBHreview}
We start with the five-dimensional Einstein-Maxwell system 
whose  action is given by
\begin{equation}
 S=\frac{1}{16\pi G}\int d^5x \sqrt{-g}\left( R - F_{\mu\nu}F^{\mu\nu} \right),
\end{equation}
where $G,~R$ and $ F_{\mu\nu} = \partial_\mu A_{\nu}-\partial_\nu A_{\mu} $ 
are the five-dimensional gravitational constant, 
the Ricci scalar curvature 
and the Maxwell field strength with the gauge potential $A_{\mu}$.
{}From this action, we obtain the Einstein equation 
\begin{eqnarray}
 R_{\mu\nu}-\frac{1}{2} R g _{\mu\nu} 
=
2 T_{\mu \nu},\label{Einsteineq}
\end{eqnarray}
with
\begin{eqnarray}
T_{\mu \nu} = F_{\mu\lambda} F_{\nu}{}^{\lambda}
-
\frac{1}{4} g _{\mu\nu} F_{\alpha \beta} F^{\alpha \beta},
\end{eqnarray}
and the Maxwell equation
\begin{eqnarray}
\nabla_\mu F^{\mu\nu} = 0. \label{Maxwelleq}
\end{eqnarray}

SqKK black holes with charge~\cite{Ishihara:2005dp} are solutions of Eqs. (\ref{Einsteineq}) and (\ref{Maxwelleq})
whose metric and gauge potential are given by
\begin{eqnarray}
ds^2&=&-F(\rho)dt^2+\frac{K(\rho)^2}{F(\rho)}d\rho ^2+\rho ^2K(\rho)^2[(\sigma ^1)^2+(\sigma ^2)^2]
+\frac{(\rho_0+\rho_+)(\rho_0+\rho_-)}{K(\rho)^2}(\sigma^3)^2, \label{eq:met}
\\
A_{\mu}dx^\mu &=&  \frac{\sqrt{3}}{2}\frac{\sqrt{\rho_+ \rho_-}}{\rho} dt,
\end{eqnarray}
where the functions $F(\rho)$ and $K(\rho)$ are
\begin{eqnarray}
F(\rho) = \frac{(\rho-\rho_+)(\rho-\rho_-)}{\rho ^2},
\quad
K(\rho)^2= \frac{\rho+\rho_0}{\rho},
\end{eqnarray}
and $\rho_0$ and $\rho_\pm$ are constants which satisfy
\begin{eqnarray}
\rho_+ \ge \rho_- \ge  0,
\quad
\rho_-+\rho_0 \ge 0,
\label{eq:para}
\end{eqnarray}
and the invariant one forms $\sigma^a~(a=1,2,3)$ of $SU(2)$ are given by
\begin{eqnarray}
\sigma^1 =-\sin\psi d\theta+\cos\psi \sin\theta d\phi,
\quad
\sigma^2 =\cos\psi d\theta+\sin\psi \sin\theta d\phi,
\quad
\sigma^3 = d\psi+\cos\theta d\phi.
\end{eqnarray}
The domains of angular coordinates are $0\leq \theta \leq \pi,0\leq \phi \leq 2\pi,0\leq \psi \leq 4\pi$, 
and radial coordinate $\rho$ runs in the range $0<\rho <\infty $. Horizons locate at $\rho =\rho_{\pm}$.

In the region far from the horizon, 
the metric (\ref{eq:met}) becomes
\begin{eqnarray}
ds^2 \simeq -dt^2
+
d\rho^2
+
\rho^2
(d\theta^2 + \sin^2\theta d\phi^2)
+
\frac{r^2_\infty}{4}(d\psi+\cos\theta d\phi)^2
+{\cal O}(1/\rho),
\label{eq:asym}
\end{eqnarray}
where $r_\infty = 2\sqrt{(\rho_++\rho_0)(\rho_-+\rho_0)} $ is the size of the extra-dimension.
So we can see that 
the spacetime behaves effectively four-dimensional Minkowski spacetime
if we focus on physical phenomena whose typical size is much larger than $r_\infty$.
On the other hand, in the vicinity of the horizon,
the metric (\ref{eq:met}) has five-dimensional nature.
To see this we introduce new coordinates $\eta, r$
and new parameters $r_\pm$ as
\begin{eqnarray}
t = \frac{4\rho_0^2}{r_{\infty}^2} \eta,
\quad
\rho = \frac{\rho_0 r^2}{r_{\infty}^2 - r^2},
\quad
\rho_\pm = \frac{\rho_0 r_{\pm}^2}{r_{\infty}^2 - r_\pm^2},
\end{eqnarray}
then the metric (\ref{eq:met}) becomes
\begin{eqnarray}
ds^2&=&-f(r)d\eta^2+\frac{k(r)^2}{f(r)}dr^2+r^2\left[
k(r)\left[(\sigma ^1)^2+(\sigma ^2)^2\right]
+(\sigma^3)^2
\right],
\label{nearmet}
\end{eqnarray}
with
\begin{eqnarray}
f(r) = \frac{(r^2-r_+^2)(r^2-r_-^2)}{r^4},
\quad
k(r) = \frac{(r_\infty^2-r_+^2)(r_\infty^2-r_-^2)}{(r_\infty^2-r^2)^2}.
\end{eqnarray}
In this coordinate, the horizons locate at $r = r_\pm$ and
the shape of horizon is a squashed $S^3$ which is characterized by the function $k(r)$.
Especially, 
if the size of the horizon is much smaller than that of extra-dimension $r_{\infty} \gg r_{\pm}$, 
then we can see that the function $k(r) \simeq 1$ and 
the metric (\ref{nearmet}) behaves almost five-dimensional 
Reissner-Nordstr\"{o}m black holes in the vicinity of the horizon.

The Komar mass $M$ and electric charge $Q$ for this SqKK black hole are given 
by~\cite{Ishihara:2005dp,Kurita:2007hu}
\begin{eqnarray}
M = \frac{3\pi r_\infty}{4G}(\rho_+ + \rho_-),
\quad
Q = \frac{\sqrt{3}\pi r_\infty}{G}\sqrt{\rho_+\rho_-}.
\label{masscharge}
\end{eqnarray}

For later calculations, it is convenient to define the new invariant forms,
\begin{eqnarray}
 \sigma^{\pm} = \frac{1}{2}(\sigma^1 \mp i \sigma^2)\ . 
\end{eqnarray}
The metric (\ref{eq:met}) can be rewritten by using $\sigma^\pm$,
\begin{eqnarray} 
ds^2 = -F(\rho)dt^2 + \frac{K(\rho)^2}{F(\rho)}d\rho^2 + 4\rho^2
  K(\rho)^2\sigma^+ \sigma^- +
  \frac{(\rho_0+\rho_+)(\rho_0+\rho_-)}{K(\rho)^2}(\sigma^3)^2\ .
 \label{metric+-}
\end{eqnarray}

\subsection{Classification of the perturbations based on the symmetry }\label{sec:Wigner}
Next, we
introduce
the formalism to classify metric perturbations based on 
the symmetry~\cite{Hu:1974hh,Murata:2007gv,Kimura:2007cr}.
The metric (\ref{eq:met}) has spatial symmetry of $SU(2)\times U(1) \simeq U(2)$
which is 
generated by the Killing vectors
\begin{eqnarray}
 \xi_x &=& \cos\phi\partial_\theta +
 \frac{\sin\phi}{\sin\theta}\partial_\psi -
 \cot\theta\sin\phi\partial_\phi, 
\notag \\
 \xi_y &=& -\sin\phi\partial_\theta +
 \frac{\cos\phi}{\sin\theta}\partial_\psi -
 \cot\theta\cos\phi\partial_\phi,
\notag \\
 \xi_z &=& \partial_\phi, 
\notag \\
 \bm{e}_3 &=& \partial_\psi.
\end{eqnarray}
We treat the metric perturbation $g_{\mu \nu}+h_{\mu \nu}$ and the electromagnetic perturbation $A_\mu+\delta A_\mu$,
where $g_{\mu \nu}$ and $A_\mu$ are the background quantities.
Because of the background symmetry, 
these perturbations can be expanded by using
Wigner functions which are known as the irreducible representation of 
$SU(2)\times U(1)$~\cite{Hu:1974hh,Murata:2007gv,Kimura:2007cr}.
Scalar Wigner function is defined by
\begin{eqnarray}
 L^2 D^J_{KM} = J(J+1)D^J_{KM}\ ,  \quad
 L_z D^J_{KM} = M D^J_{KM}\ ,  \quad
 W_3 D^J_{KM} = K D^J_{KM}\ ,\label{eq:WigDef}
\end{eqnarray}
where 
$L_\alpha := i \xi_\alpha~(\alpha = x,y,z),~W_3 := i \bm{e}_3,~L^2 := L_x^2 + L_y^2 + L_z^2$, 
and $J,K,M$ are integers or half integers\footnote{
$K, M$ take integers iff $J$ takes integers.}
satisfying $J\geq 0,\  |K|\leq J,\  |M|\leq J$. 
Vector Wigner functions are constructed by the action of invariant forms $\sigma^a$ on scalar Wigner functions,
\begin{eqnarray}
D_{i,K}^+ &=& \sigma^+_i D_{K-1}, \quad (|K-1|\leq J),
\notag \\
D_{i,K}^- &=& \sigma^-_i D_{K+1}, \quad (|K+1|\leq J),
\notag \\
D_{i,K}^3 &=& \sigma^3_i D_{K}, \qquad (|K|\leq J).
\label{basis-v}
\end{eqnarray}
Here, we have omitted the subscript $J,M$. 
The vector Wigner functions
satisfy
\begin{eqnarray}
 L^2 D_{i,K}^a = J(J+1)D_{i,K}^a,\quad
 L_z D_{i,K}^a = M D_{i,K}^a,\quad
 W_3 D_{i,K}^a = K D_{i,K}^a,
\label{eq:vec_Wigner}
\end{eqnarray}
where $a=\pm,3$ and operations are defined by Lie derivatives; 
$W_3 D_{i,K}^b := \mathcal{L}_{W_3} D_{i,K}^b$ and 
$L_\alpha D_{i,K}^a := \mathcal{L}_{L_\alpha} D_{i,K}^a$. 
Tensor Wigner functions $D_{ij,K}^{ab}$ are defined by
\begin{eqnarray}
D_{ij,K}^{++} &=& \sigma^+_i \sigma^+_j D_{K-2} \quad(|K-2|\leq J),
\notag \\
D_{ij,K}^{+-} &=& \sigma^+_i \sigma^-_j D_{K} \qquad(|K|\leq J),
\notag \\
D_{ij,K}^{+3} &=& \sigma^+_i \sigma^3_j D_{K-1} \quad(|K-1|\leq J),
\notag \\
D_{ij,K}^{--} &=& \sigma^-_i \sigma^-_j D_{K+2} \quad(|K+2|\leq J),
\notag \\
D_{ij,K}^{-3} &=& \sigma^-_i \sigma^3_j D_{K+1} \quad(|K+1|\leq J),
\notag \\
D_{ij,K}^{33} &=& \sigma^3_i \sigma^3_j D_{K} \qquad(|K|\leq J).
\label{basis-t}
\end{eqnarray}
One can check that $D_{ij,K}^{ab}$ forms the irreducible representation:
\begin{eqnarray}
 L^2 D_{ij,K}^{ab} = J(J+1)D_{ij,K}^{ab},
\quad
 L_z D_{ij,K}^{ab} = M D_{ij,K}^{ab},
\quad
 W_3 D_{ij,K}^{ab} = K D_{ij,K}^{ab}. 
\label{eq:tensor_Wigner}
\end{eqnarray}
The tensor field $h_{\mu\nu}$ can be divided into three parts, 
$h_{AB},h_{Ai},h_{ij},~(A,B = t, \rho)$ which behave as scalar, vector and tensor 
within the submanifold $\theta,\phi,\psi$.
Similarly, the vector field $\delta A_{\mu}$ can be divided into two parts, 
$\delta A_{A}$ and $\delta A_{i}$.
These perturbations can be expanded using the Winger functions as 
\begin{eqnarray}
h_{\mu \nu}dx^\mu dx^\nu &=& dx^Adx^B \sum_K h_{AB}^K(x^A)D_K(x^i) + 2dx^Adx^i \sum_K h_{Aa}^K(x^A)D^a_{i,K}(x^i)
\notag \\
                         & & +dx^idx^j\sum_K h_{ab}^K(x^A)D^{ab}_{ij,K}(x^i),
\notag \\
\delta A_\mu dx^\mu      
&=& dx^A\sum_K \delta A_A^K(x^A)D_K(x^i)+dx^i\sum_K \delta A_a(x^A)D^a_{i,K}(x^i).
\end{eqnarray}
Because perturbed quantities are expanded in terms of the representation of
the spacetime symmetry, no coupling appears
between coefficients with different sets of indices $(J,M,K)$ in the perturbed equations.
In addition, we utilize the Fourier expansion with respect to the time coordinate $t$.

Interestingly, without explicit calculation, we can 
reveal the structure of couplings between coefficients with the same $(J,M,K)$. 
First, since the index $K$ is shifted in the definition of 
vector and tensor harmonics, 
then the coefficients $h_{AB}^K$, $h_{Aa}^K$ and $h_{ab}^K$ exist only if
 $K$ satisfies the inequality listed in the following table:
\begin{equation}
\begin{array}{|c|c|c|c|c|}
\hline
h_{++} & h_{A+},h_{+3}, \delta A_{+} & h_{AB}, h_{A3} , h_{+-},h_{33}, \delta A_{A}, \delta A_3 
& h_{A-},h_{-3},\delta A_{-} &h_{--} \\ \hline
|K-2|\leq J &|K-1|\leq J    &|K|\leq J  &|K+1|\leq J   &|K+2|\leq J \\ \hline
\end{array}
\nonumber
\end{equation}
Therefore, for zero mode $J=0$, we can classify the coefficients by possible $K$ as follows:

\hspace{1cm} $J=0;$
\begin{equation}
\begin{array}{|c|c|c|c|c|}
\hline
h_{++} & h_{A+},h_{+3}, \delta A_{+} & h_{AB}, h_{A3} , h_{+-},h_{33}, \delta A_{A}, \delta A_3 
& h_{A-},h_{-3},\delta A_{-} &h_{--} \\ \hline
K=2    &               &                        &              &       \\ \hline
       &K=1            &                        &              &       \\ \hline
       &               &K=0                     &              &       \\ \hline
       &               &                        &K=-1          &       \\ \hline
       &               &                        &              &K=-2   \\ \hline
\end{array}
\nonumber
\end{equation}
Apparently, for $h_{++}$ and $h_{--}$, we can obtain equations for a single variable, 
respectively.
For other sets of coefficients 
$(h_{A+},h_{+3},\delta A_+)$, $(h_{AB}, h_{A3} , h_{+-},h_{33},\delta A_{A}, \delta A_3)$,
$(h_{A-},h_{-3},\delta A_-)$ 
they are coupled to the coefficients in each set. 
As we will see later, after fixing the gauge symmetry,
we have the coupled master equations for one or two master variables in each set.
We can also easily see that $h_{++}$ in $(J,M,K=J+2)$ modes and
$h_{--}$ in $(J,M,K=-(J+2))$ modes are always decoupled.
The perturbed equations for these modes can be reduced to the 
master equations for the single variables, respectively.
We discuss the stability for $J=0$ modes and $K=\pm (J+2)$ modes in the next section.

\section{Stability analysis}\label{sec:stability}
The linearized Einstein equation is
\begin{eqnarray}
\delta G_{\mu \nu}-2 \delta T_{\mu \nu}=0, \label{eq:einstein}
\end{eqnarray}
where the linearized Einstein tensor $\delta G_{\mu \nu}$ and the linearized stress tensor $\delta T_{\mu \nu}$ are given by
\begin{eqnarray}
\delta G_{\mu \nu}
&=&
\frac{1}{2}[
\nabla ^{\rho}\nabla _{\mu}h_{\nu \rho}
+
\nabla ^{\rho}\nabla _{\nu}h_{\mu \rho}
-
\nabla ^2h_{\mu \nu}
-
\nabla_{\mu}\nabla_{\nu}h 
\notag \\ &&
-g_{\mu \nu}(
  \nabla^{\rho}\nabla^{\lambda}h_{\rho \lambda}
  -
  \nabla^{\rho}\nabla_{\rho}h
  +
  h^{\rho \lambda}R_{\rho \lambda} )
-
h_{\mu \nu}R],
\\
\delta T_{\mu \nu}
&=&
-
h_{\alpha \beta}F_\mu{}^\alpha F_\nu{}^\beta
-
\frac{1}{4}h_{\mu \nu}F_{\alpha \beta}F^{\alpha \beta}
+
\frac{1}{2}g_{\mu \nu}h_{\alpha \beta}F^{\alpha \rho}F^\beta{}_\rho
+
\delta F_{\mu \alpha}F_\nu{}^\alpha
\notag \\ &&
+
\delta F_{\nu \alpha}F_\mu{}^\alpha
-
\frac{1}{2}g_{\mu \nu}\delta F_{\alpha \beta}F^{\alpha \beta},
\end{eqnarray}
where
\begin{eqnarray}
h=g^{\mu \nu}h_{\mu \nu},
\quad
\delta F_{\mu \nu} = \partial _{\mu}(\delta A_{\nu})-\partial _{\nu}(\delta A_{\mu}),
\end{eqnarray}
and $R_{\mu \nu},F_{\mu \nu}$ are the background quantities.
Note that $\nabla _{\mu}$ denotes the covariant derivative with respect to the background metric $g_{\mu \nu}$.
The Maxwell equation for the perturbations is
\begin{eqnarray}
\delta(\nabla _{\mu}F^{\mu \nu})
&=&
g^{\nu \beta}\nabla^\alpha \delta F_{\alpha \beta}
-
\nabla^\beta (h_{\beta \alpha}F^{\alpha \nu})
-
g^{\nu \alpha}\nabla_\mu (h_{\alpha \beta}F^{\mu \beta})
+\frac{1}{2}F^{\alpha \nu}\nabla_\alpha h-\frac{1}{2}F^{\alpha \beta}\nabla ^\nu h_{\alpha \beta}
\notag\\ 
&=&
0. \label{eq:maxwell}
\end{eqnarray}

We treat two kind of gauge transformation. One is related to infinitesimal coordinate transformation:
\begin{eqnarray}
h_{\mu \nu} &\to & h_{\mu \nu}+\nabla _{\mu}\xi _{\nu}+\nabla _{\nu}\xi _{\mu}, \label{eq:h,gauge}
\\
\delta A_\mu &\to & \delta A_\mu +\xi^\nu \nabla_\nu A_\mu+A_\nu \nabla_\mu \xi^\nu,
\end{eqnarray}
where $\xi_\mu$ is arbitrary vector field.
The other is related to $U(1)$ gauge:
\begin{eqnarray}
\delta A_\mu \to \delta A_\mu+\nabla_\mu \chi, \label{eq:A,gauge}
\end{eqnarray}
where $\chi$ is arbitrary scalar field.

\subsection{Zero mode perturbations $(J=0,M=0)$}
In the case of $J=0,M=0$, there are five modes, $K=\pm 2,\pm 1,0$. We analyze these modes separately.
These modes correspond to perturbations with $SU(2)$ symmetry.

\subsubsection{$K=\pm 2$ modes}
\begin{figure}[htbp]
 \begin{center}
 \includegraphics[width=9cm,clip]{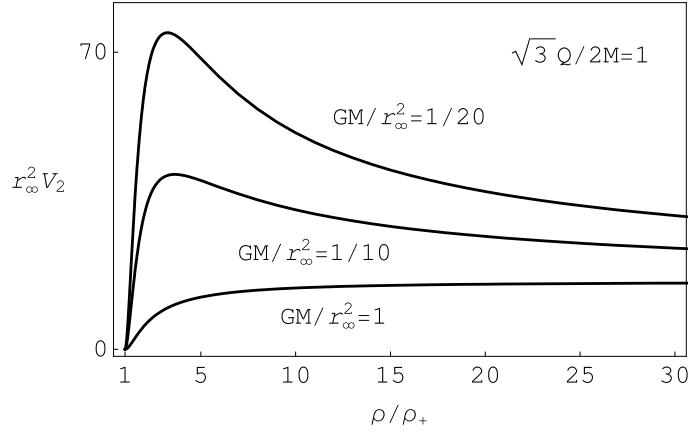}
 \end{center}
 \caption{
The effective potential $r_\infty ^2V_2$ in maximally charged case.
}
 \label{figK2;1}
\end{figure}

\begin{figure}[htbp]
 \begin{center}
 \includegraphics[width=9cm,clip]{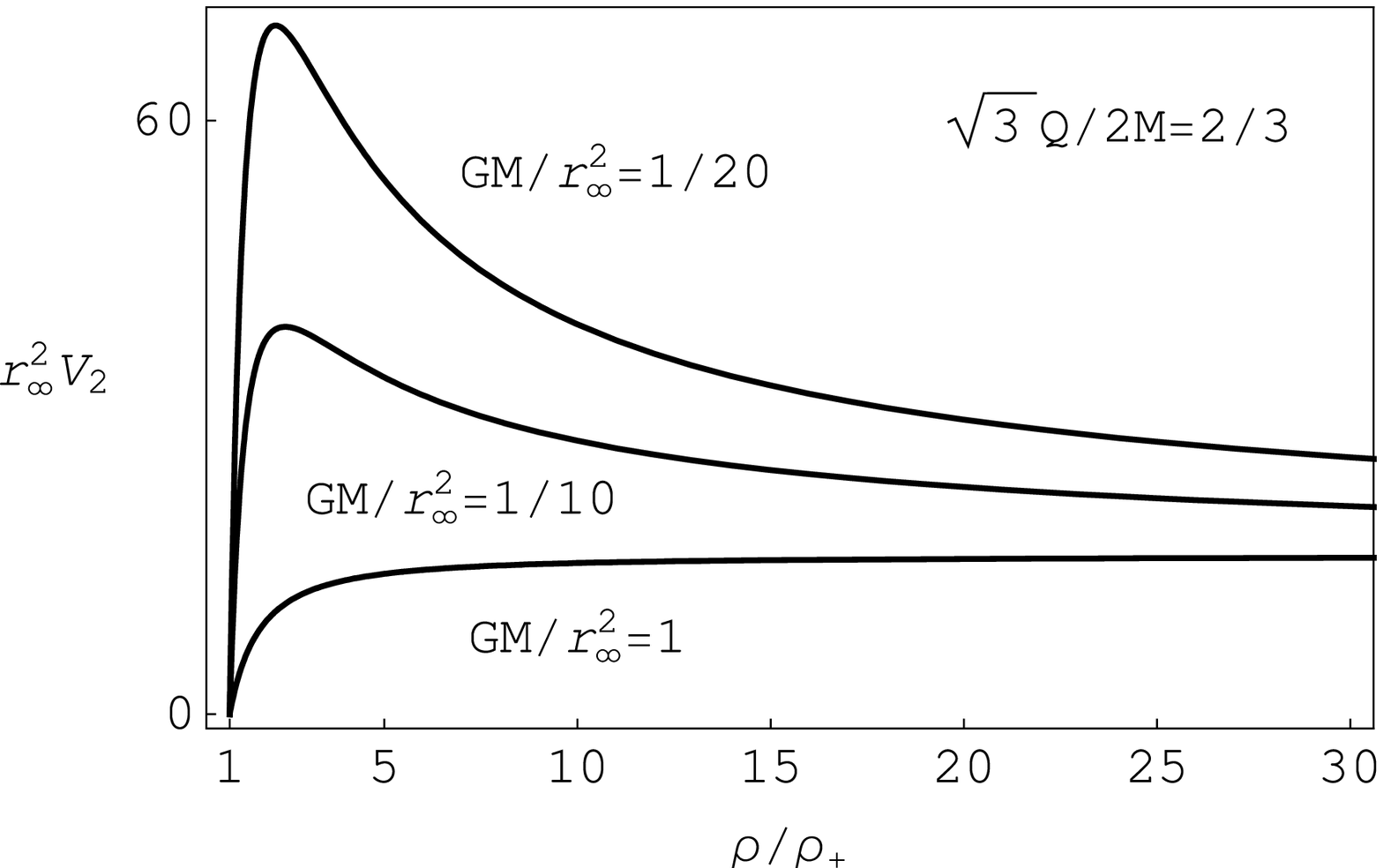}
 \end{center}
 \caption{
The effective potential $r_\infty ^2V_2 $ in $\sqrt{3}Q/2M=2/3$ case.
}
 \label{figK2;2/3}
\end{figure}

\begin{figure}[htbp]
 \begin{center}
 \includegraphics[width=9cm,clip]{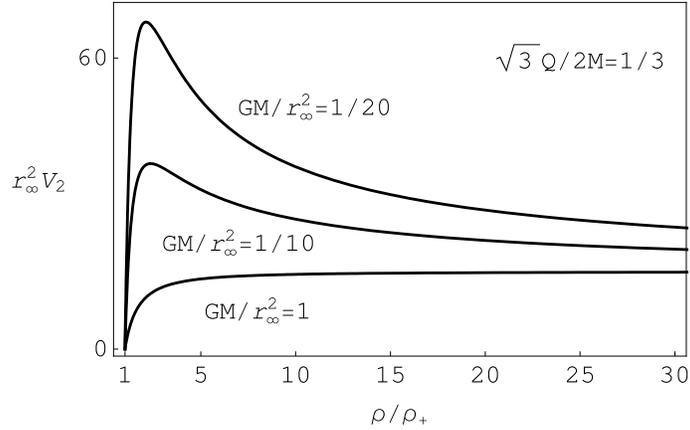}
 \end{center}
 \caption{
The effective potential $r_\infty ^2V_2 $ in $\sqrt{3}Q/2M=1/3$ case.
}
 \label{figK2;1/3}
\end{figure}

In $K=\pm 2$ modes, there are two coefficients $h_{++}$ and $h_{--}$, and these are gauge invariant. 
Because the perturbed metric $h_{\mu \nu}(x^{\mu})dx^{\mu}dx^{\nu}$ is real, 
$h_{--}$ must be the complex conjugate of 
$h_{++}$, i.e. $\bar{h}_{++}=h_{--}$ where the bar denotes the
complex conjugate.
So, it is sufficient to treat only $h_{++}$. 
We set $h_{\mu \nu}$ as
\begin{eqnarray}
h_{\mu \nu}(x^{\mu})dx^{\mu}dx^{\nu}=h_{++}(\rho)e^{-i\omega t}\sigma^+ \sigma^+.
\label{eq:K2,met}
\end{eqnarray}
Inserting Eq.~(\ref{eq:K2,met}) into $(++)$ component of Eq.~(\ref{eq:einstein}),
we obtain the perturbation equation for $h_{++}$ whose explicit form is written in Appendix \ref{appendixA}.
By introducing the master variable
\begin{eqnarray}
\Phi_2(\rho):= \frac{1}{\rho^{1/4}(\rho+\rho_0)^{3/4}}h_{++}(\rho),
\end{eqnarray}
and switching to the tortoise coordinate $\rho_{*}$ defined by
\begin{eqnarray}
\frac{d\rho_*}{d\rho}=\frac{K(\rho)}{F(\rho)},
\end{eqnarray}
we get the perturbation equation in the Schr\"odinger form,
\begin{eqnarray}
-\frac{d^2}{d\rho_*^2}\Phi_2+V_2(\rho)\Phi_2=\omega^2\Phi_2, \label{eq:K2,sch}
\end{eqnarray}
where the potential $V_2(\rho)$ is defined by
\begin{eqnarray}
V_2(\rho) &=& 
\frac{(\rho-\rho_+) (\rho-\rho_++\tilde{\rho}_-)}{16 \tilde{\rho}_0 (\tilde{\rho}_0+\tilde{\rho}_-) \rho^5 (\tilde{\rho}_0+\tilde{\rho}_-+\rho-\rho_+)^3}
\notag \\ & &
\times \bigl[4 \rho_+ (3 \tilde{\rho}_-+8 \rho_+) \tilde{\rho}_0^4+4 \tilde{\rho}_- \rho_+ (9 \tilde{\rho}_- 
+41 \rho_+) \tilde{\rho}_0^3+4 \tilde{\rho}_-^2 \rho_+ (9 \tilde{\rho}_-+74 \rho_+) \tilde{\rho}_0^2
\notag \\ & & 
+12 \tilde{\rho}_-^3 \rho_+ (\tilde{\rho}_-+19 \rho_+) \tilde{\rho}_0+64 \tilde{\rho}_-^4 \rho_+^2
+((88 \rho_+ -9 \tilde{\rho}_-) \tilde{\rho}_0^4
\notag \\ & & 
+(-27 \tilde{\rho}_-^2+418 \rho_+ \tilde{\rho}_-+200 \rho_+^2) \tilde{\rho}_0^3+\tilde{\rho}_- (-27 \tilde{\rho}_-^2+700 \rho_+ \tilde{\rho}_-+655 \rho_+^2) \tilde{\rho}_0^2  
\notag \\ & & 
+(-9 \tilde{\rho}_-^4+498 \rho_+ \tilde{\rho}_-^3+711 \rho_+^2 \tilde{\rho}_-^2) \tilde{\rho}_0+128 \tilde{\rho}_-^3 \rho_+ (\tilde{\rho}_-+2 \rho_+))(\rho-\rho_+) 
\notag \\ & & 
+(35 \tilde{\rho}_0^4+(149 \tilde{\rho}_-+442 \rho_+) \tilde{\rho}_0^3+(257 \tilde{\rho}_-^2+1400 \rho_+ \tilde{\rho}_-+355 \rho_+^2) \tilde{\rho}_0^2 
\notag \\ & & 
+\tilde{\rho}_- (207 \tilde{\rho}_-^2+1470 \rho_+ \tilde{\rho}_-+739 \rho_+^2) \tilde{\rho}_0+64 \tilde{\rho}_-^2 (\tilde{\rho}_-^2+8 \rho_+ \tilde{\rho}_-+6 \rho_+^2))(\rho-\rho_+)^2 
\notag \\ & & 
+(200 \tilde{\rho}_0^3+8 (80 \tilde{\rho}_-+91 \rho_+) \tilde{\rho}_0^2+8 (87 \tilde{\rho}_-^2+187 \rho_+ \tilde{\rho}_-+32 \rho_+^2) \tilde{\rho}_0
\notag \\ & & 
+256 \tilde{\rho}_- (\tilde{\rho}_-^2+3 \rho_+ \tilde{\rho}_- 
+\rho_+^2))(\rho-\rho_+)^3
\notag \\ & & 
+(352 \tilde{\rho}_0^2+(736 \tilde{\rho}_-+512 \rho_+) \tilde{\rho}_0+64 (6 \tilde{\rho}_-^2+8 \rho_+ \tilde{\rho}_-+\rho_+^2)) (\rho-\rho_+)^4 
\notag \\ & & 
+(256 \tilde{\rho}_0+128 (2 \tilde{\rho}_-+\rho_+)) (\rho-\rho_+)^5+64 (\rho-\rho_+)^6\bigr]. 
\label{eq:V2}
\end{eqnarray}
Here, we introduced new parameters $\tilde{\rho}_-$ and $\tilde{\rho}_0$ defined by
\begin{eqnarray}
\tilde{\rho}_- := \rho_+-\rho_-,\quad \tilde{\rho}_0 := \rho_0+\rho_-. \label{tilde}
\end{eqnarray}
From Eq.~(\ref{eq:para}), these parameters satisfy the following inequalities,
\begin{eqnarray}
0\leq \tilde{\rho}_-\leq \rho_+,\quad \tilde{\rho}_0>0. \label{ineq:tilde}
\end{eqnarray}
Since the terms in square bracket in Eq.~(\ref{eq:V2}) are given in a power series expansion of $(\rho -\rho_+)$ 
and the expansion coefficients of these terms are positive,
we can see $V_2>0$ in the region $\rho_+<\rho<\infty $.
Typical profiles of the potential $V_2$ are plotted in Figs.\ref{figK2;1}-\ref{figK2;1/3}.
One should be noted that we have equality $\sqrt{3}Q/(2M)=1$ in maximally charged case $\rho_+ = \rho_-$
from (\ref{masscharge}),
and we normalized the depicted effective potential by using the size of extra dimension.
The desired solution for $\Phi_2$ is square integrable in the domain $-\infty <\rho_{*} <\infty$ with real $\omega ^2$.
Multiplying both sides of Eq. (\ref{eq:K2,sch}) by $\bar{\Phi}_2$, and integrating it, we obtain
\begin{eqnarray}
 \int d\rho _*\left[ \left| \frac{d\Phi_2}{d\rho _*}\right|^2+ V_2|\Phi_2|^2 \right]
-\left[ \bar{\Phi}_2\frac{d}{d\rho_*}\Phi_2 \right] ^{\infty}_{-\infty}=\omega ^2\int d\rho_*\left| \Phi_2\right|^2.
\end{eqnarray}
Because the boundary term vanishes, the positivity of $V_2$ means $\omega ^2>0$. Therefore, We conclude that the background spacetime is stable 
with respect to the $K=\pm 2$ perturbations.

\subsubsection{$K=\pm 1$ modes}
Because of the relations $\bar{h}_{A+}=h_{A-}$, $\bar{h}_{+3}=h_{-3}$ and $\delta \bar{A}_+=\delta A_-$, 
it is sufficient to consider only $h_{A+}$, $h_{3+}$ and $\delta A_+$.
Thus, we assume
\begin{eqnarray}
h_{\mu \nu}dx^{\mu}dx^{\nu}&=&2h_{A+}(\rho)e^{-i\omega t}dx^A\sigma^+ +2h_{+3}(\rho)e^{-i\omega t}\sigma^+\sigma^3, \label{eq:K1,met}
\end{eqnarray}
\begin{eqnarray}
\delta A_{\mu}dx^{\mu}=\delta A_+(\rho)e^{-i\omega t}\sigma^+. \label{eq:K1,A}
\end{eqnarray}
Here, we set the gauge $\xi_{\mu}dx^{\mu}$ as
\begin{eqnarray}
\xi_{\mu}dx^{\mu}=\xi_+(\rho)e^{-i\omega t}\sigma^+,
\end{eqnarray}
and under the gauge transformations (\ref{eq:h,gauge}), the metric perturbations transform as
\begin{eqnarray}
h_{t+} 
&\to & 
h_{t+}-i\omega \xi_+,
\\
h_{\rho +} 
&\to &  
h_{\rho +}-\frac{2\rho+\rho_0}{\rho(\rho+\rho_0)}\xi_++\frac{d\xi_+}{d\rho},
\\
h_{3+} 
&\to &
h_{3+}-\frac{i \left( \rho^2+2 \rho_0 \rho-\rho_- \rho_+-(\rho_-+\rho_+) \rho_0\right)}{(\rho+\rho_0)^2}\xi_+.
\end{eqnarray}
So we can impose the gauge condition
\begin{eqnarray}
h_{3+}=0, \label{eq:K1,gauge}
\end{eqnarray}
which completely fixes the gauge freedom.
Substituting Eqs. (\ref{eq:K1,met}), (\ref{eq:K1,A}) and (\ref{eq:K1,gauge}) 
into Eqs. (\ref{eq:einstein}) and (\ref{eq:maxwell}),
we obtain perturbation equations whose explicit forms are given in Appendix \ref{appendixA}.
Eliminating $h_{t+}$ from these equations,
and defining new variables
\begin{eqnarray}
\phi_{1G} := \frac{(\rho -\rho_+)(\rho -\rho_-)(\rho_+\rho_-+\rho_0(\rho_++\rho_--2\rho)-\rho^2)}{\rho ^{7/4}(\rho +\rho_0)^{9/4}}h_{\rho +},
\end{eqnarray}
\begin{eqnarray}
\phi_{1E} := \frac{\rho ^{1/4}}{(\rho +\rho_0)^{1/4}}\frac{\delta A_+}{i\omega },
\end{eqnarray}
we have the coupled master equations for the gravitational and the electromagnetic perturbations
\begin{eqnarray}
-\frac{d^2}{d\rho_*^2}\phi_{1G}+V_{11}(\rho)\phi_{1G}+V_{12}(\rho)\phi_{1E}=\omega^2\phi_{1G}, \label{master1G}
\end{eqnarray}
\begin{eqnarray}
-\frac{d^2}{d\rho_*^2}\phi_{1E}+V_{22}(\rho)\phi_{1E}+V_{21}(\rho)\phi_{1G}=\omega^2\phi_{1E}, \label{master1E}
\end{eqnarray}
where
the effective potentials $V_{11}$, $V_{12}$, $V_{21}$ and $V_{22}$ are defined by
\begin{eqnarray}
V_{11} &=& 
\frac{(\rho-\rho_-)(\rho-\rho_+)}{16 \rho^5 (\rho_-+\rho_0) (\rho_++\rho_0)(\rho+\rho_0)^3 (\rho^2+2 \rho_0 \rho-\rho_- \rho_+-(\rho_-+\rho_+)\rho_0)^2}
\notag \\ & &
\times \bigl[16 \rho^{10}+128 \rho_0 \rho^9+32 (\rho_+ (\rho_-+\rho_0)+\rho_0 (\rho_-+15 \rho_0)) \rho^8
\notag \\ & & 
+8(119 \rho_0^3 
-9 (\rho_-+\rho_+) \rho_0^2-(16 \rho_-^2+25\rho_+ \rho_-+16 \rho_+^2) \rho_0-16 \rho_- \rho_+ (\rho_-+\rho_+))\rho^7
\notag \\ & & 
+(1103 \rho_0^4 
-501 (\rho_-+\rho_+) \rho_0^3-(484\rho_-^2+697 \rho_+ \rho_-+484 \rho_+^2) \rho_0^2
\notag \\ & & 
-196 \rho_- \rho_+(\rho_-+\rho_+) \rho_0 
+288 \rho_-^2 \rho_+^2) \rho^6
\notag \\ & & 
+\rho_0 (764\rho_0^4-959 (\rho_-+\rho_+) \rho_0^3-(827 \rho_-^2+634 \rho_+\rho_-+827 \rho_+^2) \rho_0^2 
\notag \\ & & 
+325 \rho_- \rho_+ (\rho_-+\rho_+) \rho_0+1152\rho_-^2 \rho_+^2) \rho^5
\notag \\ & & 
+(252 \rho_0^6-990 (\rho_-+\rho_+)\rho_0^5 
+(-722 \rho_-^2+193 \rho_+ \rho_--722 \rho_+^2)\rho_0^4
\notag \\ & & 
+(\rho_-+\rho_+) (72 \rho_-^2+1159 \rho_+ \rho_-+72\rho_+^2) \rho_0^3 
\notag \\ & & 
+3 \rho_- \rho_+ (16 \rho_-^2+611 \rho_+\rho_-+16 \rho_+^2) \rho_0^2-120 \rho_-^2 \rho_+^2 (\rho_-+\rho_+) \rho_0 
-96\rho_-^3 \rho_+^3) \rho^4
\notag \\ & & 
-2 \rho_0 (228 (\rho_-+\rho_+)\rho_0^5+(53 \rho_-^2-550 \rho_+ \rho_-+53 \rho_+^2)\rho_0^4 
\notag \\ & & 
-(\rho_-+\rho_+) (111 \rho_-^2+709 \rho_+ \rho_-+111\rho_+^2) \rho_0^3-6 \rho_- \rho_+ (7 \rho_-^2+104 \rho_+ \rho_- 
+7\rho_+^2) \rho_0^2
\notag \\ & & 
+249 \rho_-^2 \rho_+^2 (\rho_-+\rho_+) \rho_0+180 \rho_-^3\rho_+^3) \rho^3
\notag \\ & & 
+(-4 (\rho_-+\rho_0)^3 (5 \rho_0 
-4 \rho_-)\rho_+^4+\rho_0 (\rho_-+\rho_0)^2 (28 \rho_-^2-559 \rho_0 \rho_-+151\rho_0^2) \rho_+^3
\notag \\ & & 
+\rho_0^2 (\rho_-+\rho_0) (-12 \rho_-^3 
-927\rho_0 \rho_-^2+450 \rho_0^2 \rho_-+187 \rho_0^3) \rho_+^2
\notag \\ & & 
+\rho_- \rho_0^3(\rho_-+\rho_0) (-44 \rho_-^2-213 \rho_0 \rho_-+850 \rho_0^2)\rho_+ 
\notag \\ & & 
+\rho_-^2 \rho_0^4 (-20 \rho_-^2+151 \rho_0 \rho_-+187\rho_0^2)) \rho^2
\notag \\ & & 
-3 \rho_0 (\rho_-+\rho_0) (\rho_++\rho_0) 
(\rho_-\rho_++(\rho_-+\rho_+) \rho_0) (-16 \rho_-^2 \rho_+^2-7 \rho_- (\rho_-+\rho_+) \rho_0\rho_+
\notag \\ & & 
+3 (3 \rho_-^2+34 \rho_+ \rho_- 
+3 \rho_+^2)\rho_0^2) \rho
\notag \\ & & 
+39 \rho_- \rho_+ \rho_0^2 (\rho_-+\rho_0) (\rho_++\rho_0) (\rho_-\rho_++(\rho_-+\rho_+) \rho_0)^2\bigr], 
\label{V11}
\end{eqnarray}
\begin{eqnarray}
V_{12}=\frac{2 \sqrt{3} \sqrt{\rho_- \rho_+} (\rho-\rho_-) (\rho-\rho_+)
   \left(\rho^2+2 \rho_0 \rho-\rho_- \rho_+-(\rho_-+\rho_+) \rho_0\right)}{\rho^4
   (\rho+\rho_0)^2}, \label{V12}
\end{eqnarray}
\begin{eqnarray}
V_{21}=\frac{\sqrt{3} \sqrt{\rho_- \rho_+} (\rho-\rho_-) (\rho-\rho_+)
   \left(\rho^2+2 \rho_0 \rho-\rho_- \rho_+-(\rho_-+\rho_+) \rho_0\right)}{2 \rho^4
   (\rho_-+\rho_0) (\rho_++\rho_0) (\rho+\rho_0)^2}, \label{V21}
\end{eqnarray}
\begin{eqnarray}
V_{22}&=&
\frac{(\rho-\rho_-) (\rho-\rho_+)}{16 \rho^5 (\rho_-+\rho_0) (\rho_++\rho_0) (\rho+\rho_0)^3}
\bigl[16 \rho^6+64 \rho_0 \rho^5+96\rho_0^2 \rho^4 
\notag \\ & & 
+8 \rho_0 (7 \rho_0^2-(\rho_-+\rho_+) \rho_0-\rho_- \rho_+)\rho^3
\notag \\ & & 
+(15 \rho_0^4+11 (\rho_-+\rho_+) \rho_0^3+(12\rho_-^2+71 \rho_+ \rho_- 
+12 \rho_+^2) \rho_0^2
\notag \\ & & 
+60 \rho_- \rho_+ (\rho_-+\rho_+)\rho_0+48 \rho_-^2 \rho_+^2) \rho^2
\notag \\ & & 
+5 \rho_0 (\rho_-+\rho_0) (\rho_++\rho_0) 
(16 \rho_- \rho_++(\rho_-+\rho_+) \rho_0) \rho
\notag \\ & & 
+39 \rho_- \rho_+ \rho_0^2 (\rho_-+\rho_0)(\rho_++\rho_0)\bigr]. 
\label{V22}
\end{eqnarray}

In order to discuss the stability for $K=\pm 1$ modes, 
we rewrite Eq.(\ref{master1G}) and Eq.(\ref{master1E}) in the following form
\begin{eqnarray}
-\frac{d^2}{d\rho^2_*}\bm{\Phi}+\bm{V}(\rho)\bm{\Phi}=\omega ^2\bm{\Phi}, \label{eq:K1,sch}
\end{eqnarray}
where $\bm{\Phi}$ and $\bm{V}$ represent the master variables and the potential given by
\begin{eqnarray}
\bm{\Phi} =\left(
\begin{array}{c}
\phi_{1G}  \\
\phi_{1E} 
\end{array}
\right),\quad 
\bm{V}=\left(
\begin{array}{cc}
V_{11} & V_{12}  \\
V_{21} & V_{22} 
\end{array}
\right).
\end{eqnarray}
Multiplying both sides of Eq.(\ref{eq:K1,sch}) by $\bm{\Phi}^{\dagger}$, and integrating it, we obtain
\begin{eqnarray}
 \int d\rho _*\left[ \left| \frac{d\bm{\Phi}}{d\rho _*}\right|^2+\bm{\Phi}^\dagger \bm{V}\bm{\Phi} \right]
-\left[ \bm{\Phi}^\dagger\frac{d}{d\rho_*}\bm{\Phi} \right] ^{\infty}_{-\infty}=\omega ^2\int d\rho_*\left| \bm{\Phi}\right|^2,
\label{eq:K1,int}
\end{eqnarray}
where $\left| \bm{\Phi}\right|^2:= \bm{\Phi}^\dagger \bm{\Phi}$.
Because the boundary term vanishes, if the second term of the integrand in the left hand side of Eq.(\ref{eq:K1,int}) is real and positive everywhere,
there is no $\omega ^2<0$ mode.

We transform the term $\bm{\Phi}^\dagger \bm{V}\bm{\Phi}$ as follows.
We diagonalize the matrix $\bm{V}$ using the unitary transformation
\begin{eqnarray}
\bm{\Phi}^\dagger \bm{V}\bm{\Phi} &=& \bm{\Phi}^\dagger \bm{Q}^{\dagger}\bm{Q}\bm{V}\bm{Q}^{\dagger}\bm{Q}\bm{\Phi} \notag \\
                                        &=& \bm{\Psi}^\dagger \bm{\tilde {V}} \bm{\Psi} \notag \\
                                        &=& \tilde{V}_{11}|\psi_1|^2+\tilde{V}_{22}|\psi_2|^2.
\label{eq:K1,posit}
\end{eqnarray}
Here, $\bm{\tilde {V}}$ and $\bm{\Psi} $ are defined by
\begin{eqnarray}
\bm{\Psi} = \bm{Q}\bm{\Phi} =\left(
\begin{array}{c}
\psi_1  \\
\psi_2 
\end{array}
\right),\quad 
\bm{\tilde {V}} = \bm{Q}\bm{V}\bm{Q}^{\dagger}= \left(
\begin{array}{cc}
\tilde{V}_{11} &0  \\
0 & \tilde{V}_{22}
\end{array}
\right).
\end{eqnarray}
From Eq.(\ref{eq:K1,int}) and Eq.(\ref{eq:K1,posit}), 
we can see that the sufficient condition for $\omega ^2>0$ is
both $\tilde{V}_{11}$ and $\tilde{V}_{22}$ are real and positive in the domain $-\infty <\rho_{*} <\infty$.
Here, $\tilde{V}_{11}$ and $\tilde{V}_{22}$ are written by using the components of matrix $\bm{V}$ as
\begin{eqnarray}
\tilde{V}_{11} &=& \frac{V_{11}+V_{22}+\sqrt{(V_{11}-V_{22})^2+4V_{12}V_{21}}}{2}, \notag \\
\tilde{V}_{22} &=& \frac{V_{11}+V_{22}-\sqrt{(V_{11}-V_{22})^2+4V_{12}V_{21}}}{2}.
\end{eqnarray}
One can easily check that both $\tilde{V}_{11}$ and $\tilde{V}_{22}$
are real everywhere.
Then, we split the condition both $\tilde{V}_{11}>0$ and $\tilde{V}_{22}>0$ to two conditions;
(i) $\det{V}(=\det{\tilde{V}})>0$ in the domain $-\infty <\rho_{*} <\infty$, and 
(ii) $\tilde{V}_{11}>0$ and $\tilde{V}_{22}>0$ at one point out side of the horizon.

First, we show the condition  (ii).
Far from the horizon, $\tilde{V}_{11}$ and $\tilde{V}_{22}$ behave as
\begin{eqnarray}
\tilde{V}_{11} 
&=&
\frac{16(\rho-\rho_-)(\rho-\rho_+)\rho^{10}}
{16 \rho^5 (\rho_-+\rho_0) (\rho_++\rho_0)(\rho+\rho_0)^3 (\rho^2+2 \rho_0 \rho-\rho_- \rho_+-(\rho_-+\rho_+)\rho_0)^2}
+{\cal O}\Bigl(\frac{1}{\rho}\Bigr),
\label{tildeV11}
\\
\tilde{V}_{22}
&=&
\frac{16(\rho-\rho_-) (\rho-\rho_+)\rho^6}{16 \rho^5 (\rho_-+\rho_0) (\rho_++\rho_0) (\rho+\rho_0)^3}
+{\cal O}\Bigl(\frac{1}{\rho}\Bigr).
\label{tildeV22}
\end{eqnarray}
Since the dominant terms of (\ref{tildeV11}) and (\ref{tildeV22}) are positive,
the condition (ii) is fulfilled at great distance.

\begin{figure}[htbp]
 \begin{center}
 \includegraphics[width=9cm,clip]{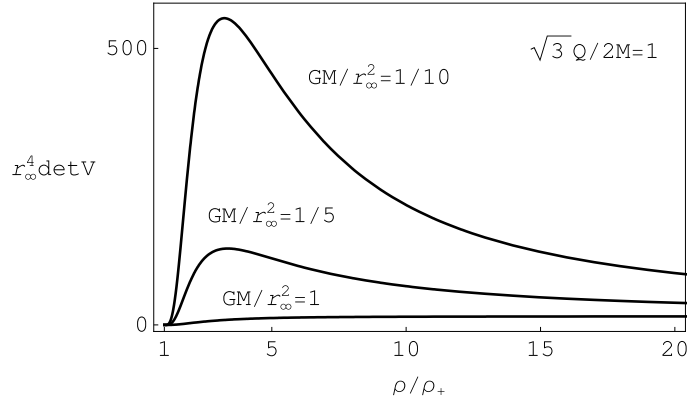}
 \end{center}
 \caption{
The determinant of the potential matrix $r_\infty ^4\det{V}$ in maximally charged case.
}
 \label{figK1;1}
\end{figure}

\begin{figure}[htbp]
 \begin{center}
 \includegraphics[width=9cm,clip]{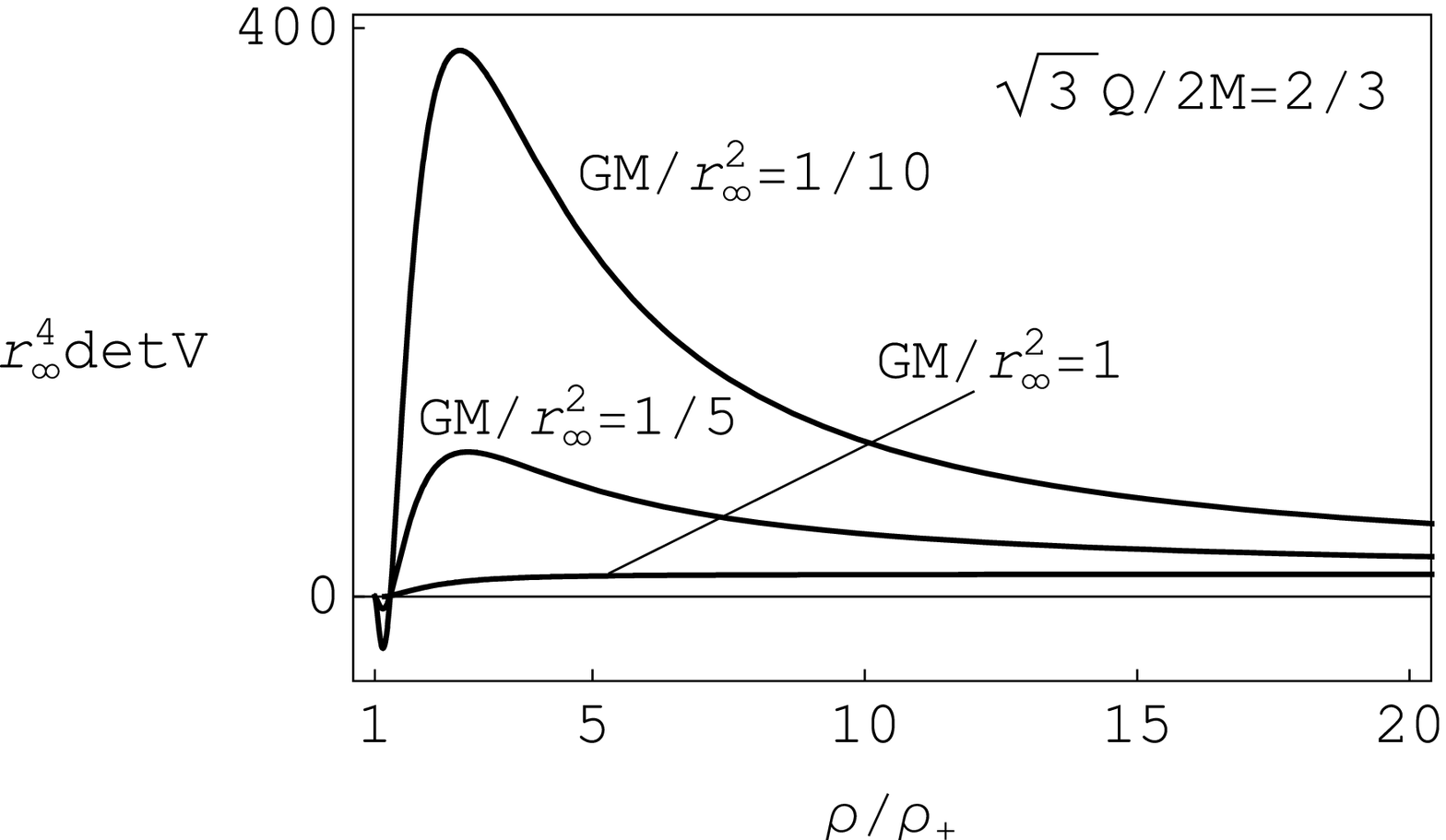}
 \end{center}
 \caption{
The determinant of the potential matrix $r_\infty ^4\det{V}$ in $\sqrt{3}Q/2M=2/3$ case.
}
 \label{figK1;2/3}
\end{figure}

\begin{figure}[htbp]
 \begin{center}
 \includegraphics[width=9cm,clip]{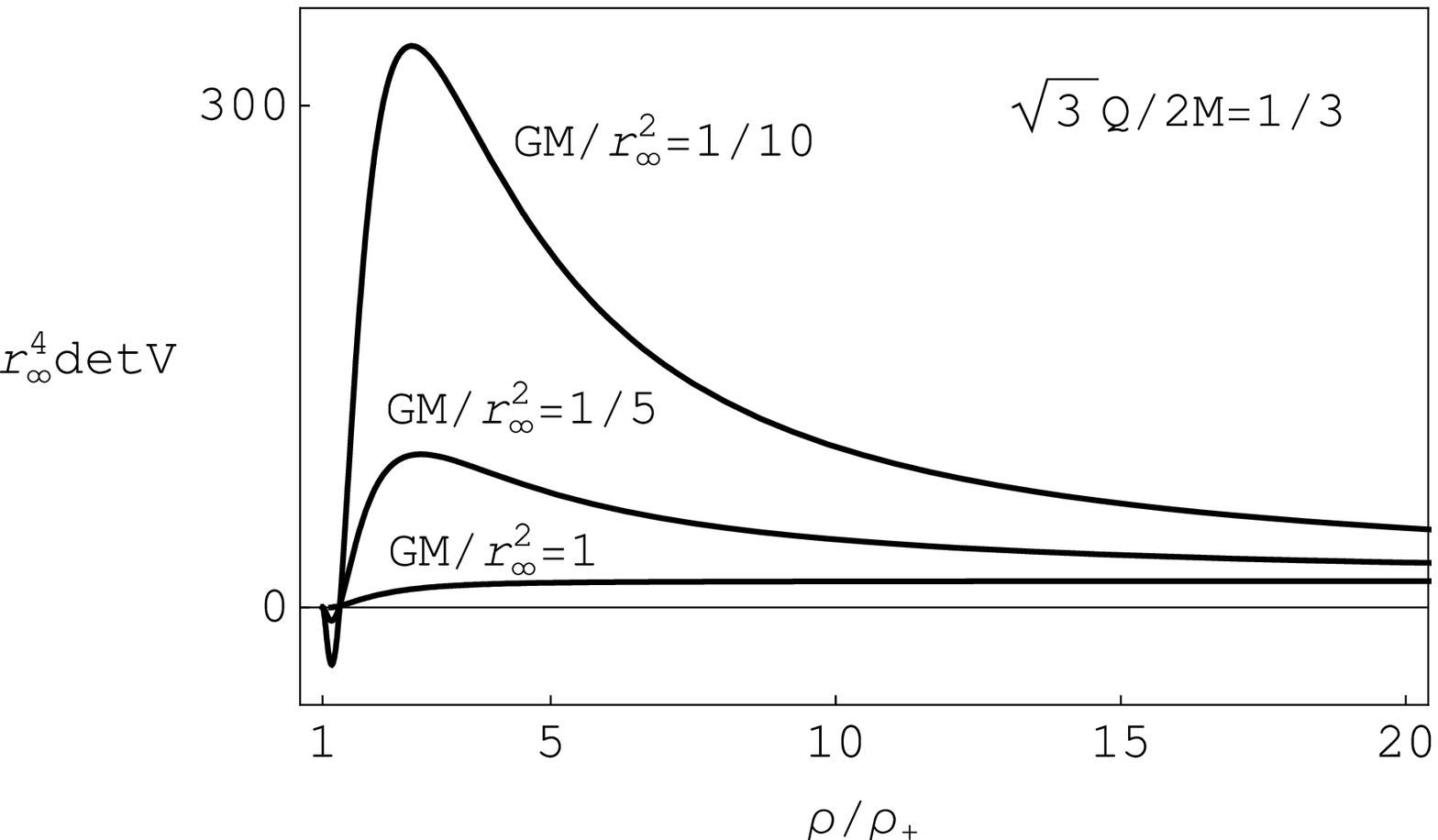}
 \end{center}
 \caption{
The determinant of the potential matrix $r_\infty ^4\det{V}$ in $\sqrt{3}Q/2M=1/3$ case.
}
 \label{figK1;1/3}
\end{figure}

Next, we consider the condition (i).
Typical profiles of $\det{V}$ are plotted in Figs.\ref{figK1;1}-\ref{figK1;1/3}.
In the case $\sqrt{3}Q /(2 M)< 1$, because there are domains in which the $\det{V}$ is negative (see Figs.\ref{figK1;2/3} and \ref{figK1;1/3}), we have not show the stability
for this case.
On the other hand, we can show the positivity of $\det{V}$ in maximally charged case
$\sqrt{3}Q /(2 M) = 1$  (see Fig.\ref{figK1;1}).
In maximally charged case, we can write $\det{V}$ as
\begin{eqnarray}
\det{V}&=&
\frac{(\rho-\rho_+)^5}{256 \tilde{\rho_0}^4 \rho^{10} (\tilde{\rho_0}-\rho_++\rho)^6 (2\tilde{\rho_0}-\rho_++\rho)^2}
\bigl[256 (\rho-\rho_+)^{13}
\notag \\ & & 
+(3072 \tilde{\rho_0}+1024 \rho_+) 
(\rho-\rho_+)^{12}+(17408 \tilde{\rho_0}^2+12288 \rho_+ \tilde{\rho_0}+1536\rho_+^2) (\rho-\rho_+)^{11}
\notag \\ & & 
+(59136\tilde{\rho_0}^3 
+67840 \rho_+ \tilde{\rho_0}^2+18432 \rho_+^2 \tilde{\rho_0}+1024\rho_+^3) (\rho-\rho_+)^{10}
\notag \\ & & 
+(132064\tilde{\rho_0}^4+219712 \rho_+ \tilde{\rho_0}^3 
+99552 \rho_+^2 \tilde{\rho_0}^2+12288 \rho_+^3\tilde{\rho_0}+256 \rho_+^4) (\rho-\rho_+)^9
\notag \\ & & 
+(203008\tilde{\rho_0}^5+459968 \rho_+ \tilde{\rho_0}^4 
+309120 \rho_+^2 \tilde{\rho_0}^3
\notag \\ & & 
+65216 \rho_+^3\tilde{\rho_0}^2+3072 \rho_+^4 \tilde{\rho_0}) (\rho-\rho_+)^8
\notag \\ & & 
+(219328\tilde{\rho_0}^6+656000 \rho_+ \tilde{\rho_0}^5 
+611232 \rho_+^2 \tilde{\rho_0}^4
\notag \\ & & 
+195648 \rho_+^3\tilde{\rho_0}^3+16096 \rho_+^4 \tilde{\rho_0}^2)(\rho-\rho_+)^7
\notag \\ & & 
+(165520 \tilde{\rho_0}^7 
+659024 \rho_+\tilde{\rho_0}^6+819120 \rho_+^2 \tilde{\rho_0}^5
\notag \\ & & 
+366448 \rho_+^3 \tilde{\rho_0}^4+47104\rho_+^4 \tilde{\rho_0}^3) (\rho-\rho_+)^6 
\notag \\& &
+(83521\tilde{\rho_0}^8+472188 \rho_+ \tilde{\rho_0}^7+788454 \rho_+^2 \tilde{\rho_0}^6
\notag \\ & & 
+449980 \rho_+^3\tilde{\rho_0}^5+84641 \rho_+^4 \tilde{\rho_0}^4) 
(\rho-\rho_+)^5
\notag \\ & & 
+(25572 \tilde{\rho_0}^9+234624 \rho_+ \tilde{\rho_0}^8+574056 \rho_+^2 \tilde{\rho_0}^7
\notag \\ & & 
+378400 \rho_+^3 \tilde{\rho_0}^6+96020 
\rho_+^4 \tilde{\rho_0}^5) (\rho-\rho_+)^4
\notag \\ & & 
+(3780\tilde{\rho_0}^{10}+73616 \rho_+ \tilde{\rho_0}^9+313560 \rho_+^2 \tilde{\rho_0}^8
\notag \\ & & 
+233104 
\rho_+^3 \tilde{\rho_0}^7+69188 \rho_+^4 \tilde{\rho_0}^6)(\rho-\rho_+)^3
\notag \\ & & 
+(11520 \rho_+ \tilde{\rho_0}^{10}+113280 \rho_+^2 
\tilde{\rho_0}^9+110848 \rho_+^3 \tilde{\rho_0}^8
+32640 \rho_+^4 \tilde{\rho_0}^7)(\rho-\rho_+)^2
\notag \\ & & 
+(19968 \rho_+^2 \tilde{\rho_0}^{10} 
+36864 \rho_+^3\tilde{\rho_0}^9+10752 \rho_+^4 \tilde{\rho_0}^8) (\rho-\rho_+)
\notag \\ & & 
+2048 \tilde{\rho_0}^9\rho_+^4+6144 \tilde{\rho_0}^{10} \rho_+^3\bigr], 
\label{detV}
\end{eqnarray}
where $\tilde{\rho_0}$ is defined by Eq.(\ref{tilde}).
{}From (\ref{ineq:tilde}) and (\ref{detV}), we can see the positivity of $\det{V}$ out side of the horizon. 
Since we have shown the condition (i) and (ii),
we conclude that the background spacetime is stable against 
perturbation for $K=\pm 1$ modes in maximally charged case.

\subsubsection{$K=0$ mode}
\begin{figure}[htbp]
 \begin{center}
 \includegraphics[width=9cm,clip]{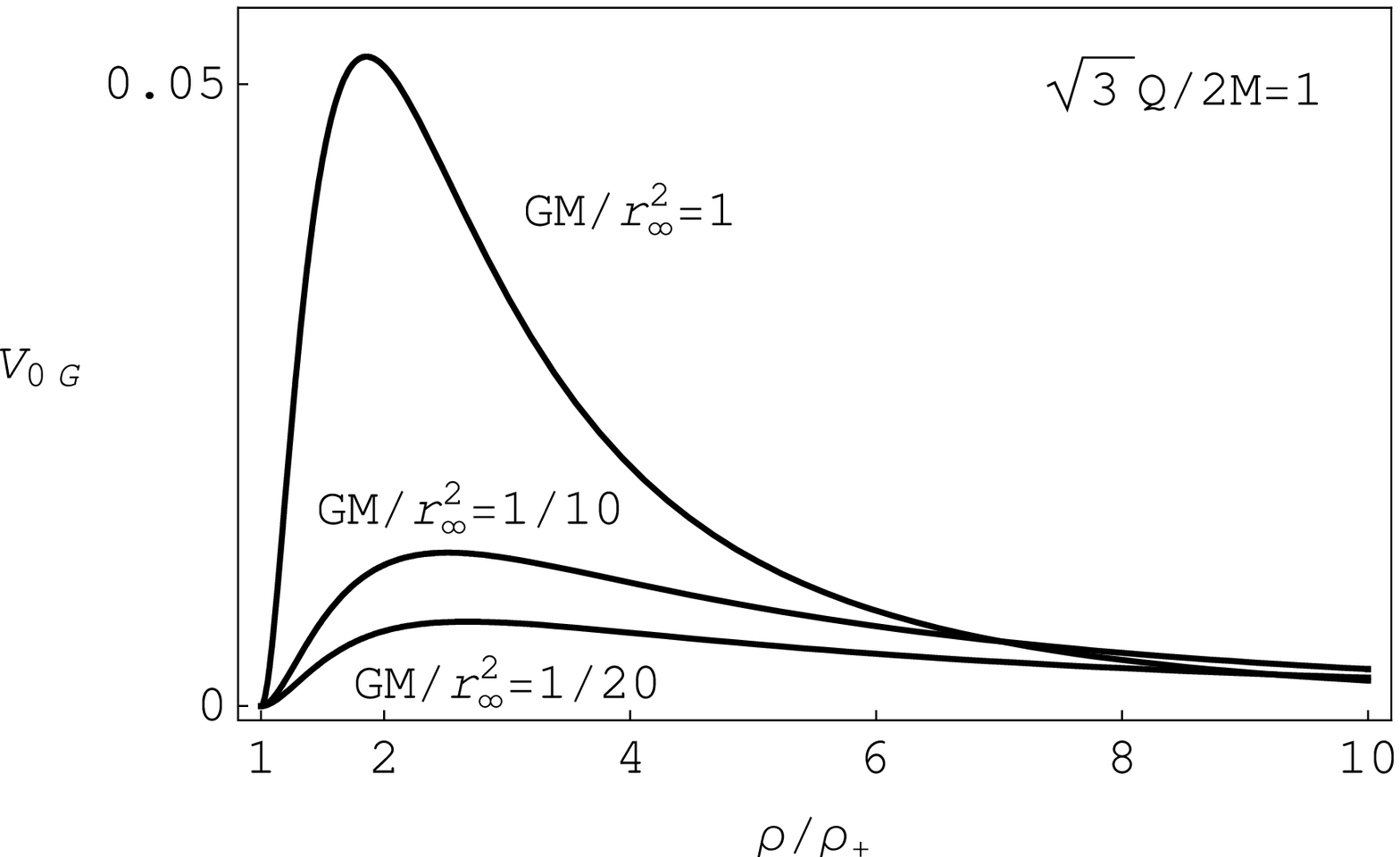}
 \end{center}
 \caption{
The effective potential $V_{0G}$ in maximally charged case.
}
 \label{figK0G;1}
\end{figure}

\begin{figure}[htbp]
 \begin{center}
 \includegraphics[width=9cm,clip]{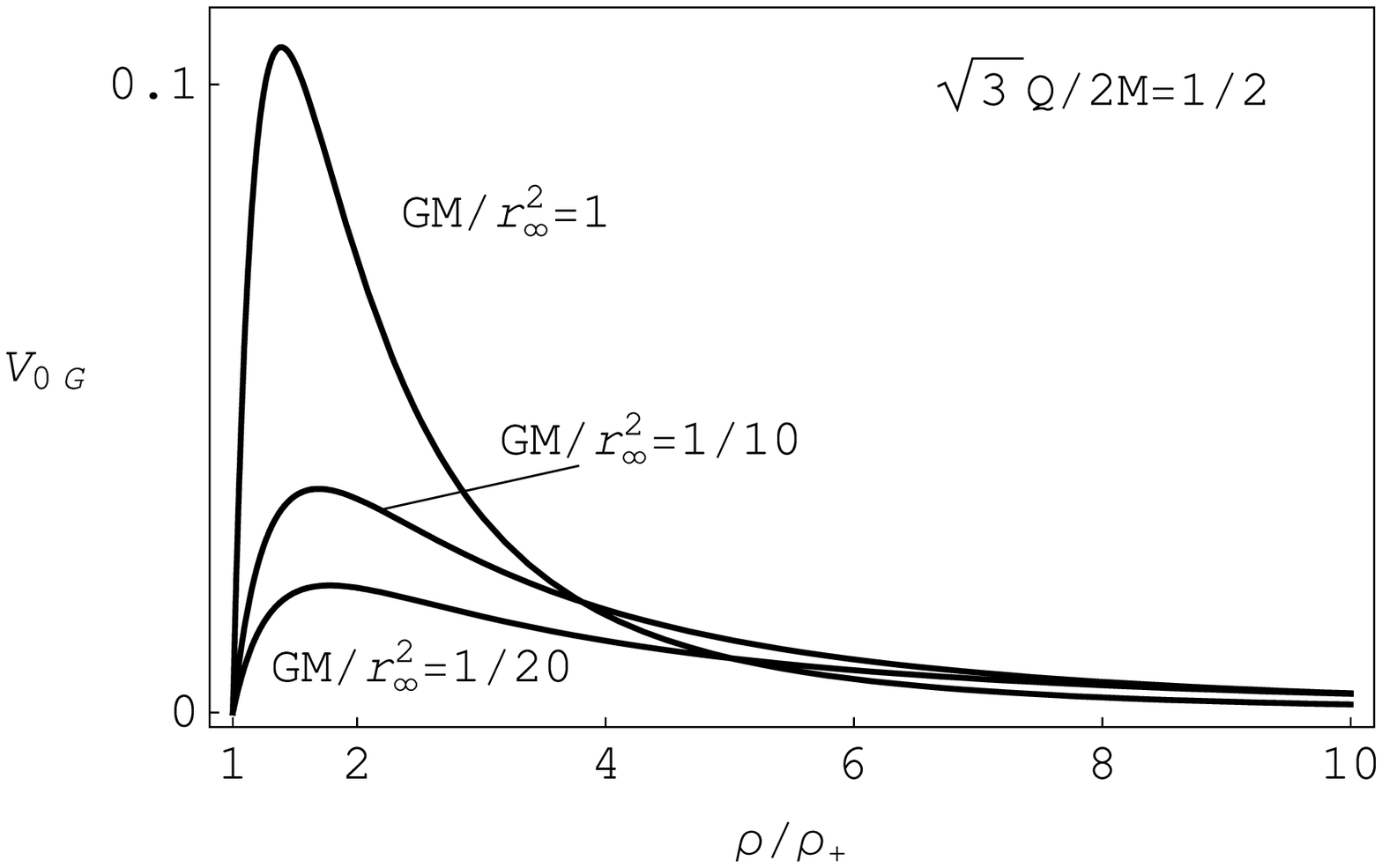}
 \end{center}
 \caption{
The effective potential $V_{0G}$ in $\sqrt{3}Q/2M=1/2$ case.
}
 \label{figK0G;1/2}
\end{figure}

\begin{figure}[htbp]
 \begin{center}
 \includegraphics[width=9cm,clip]{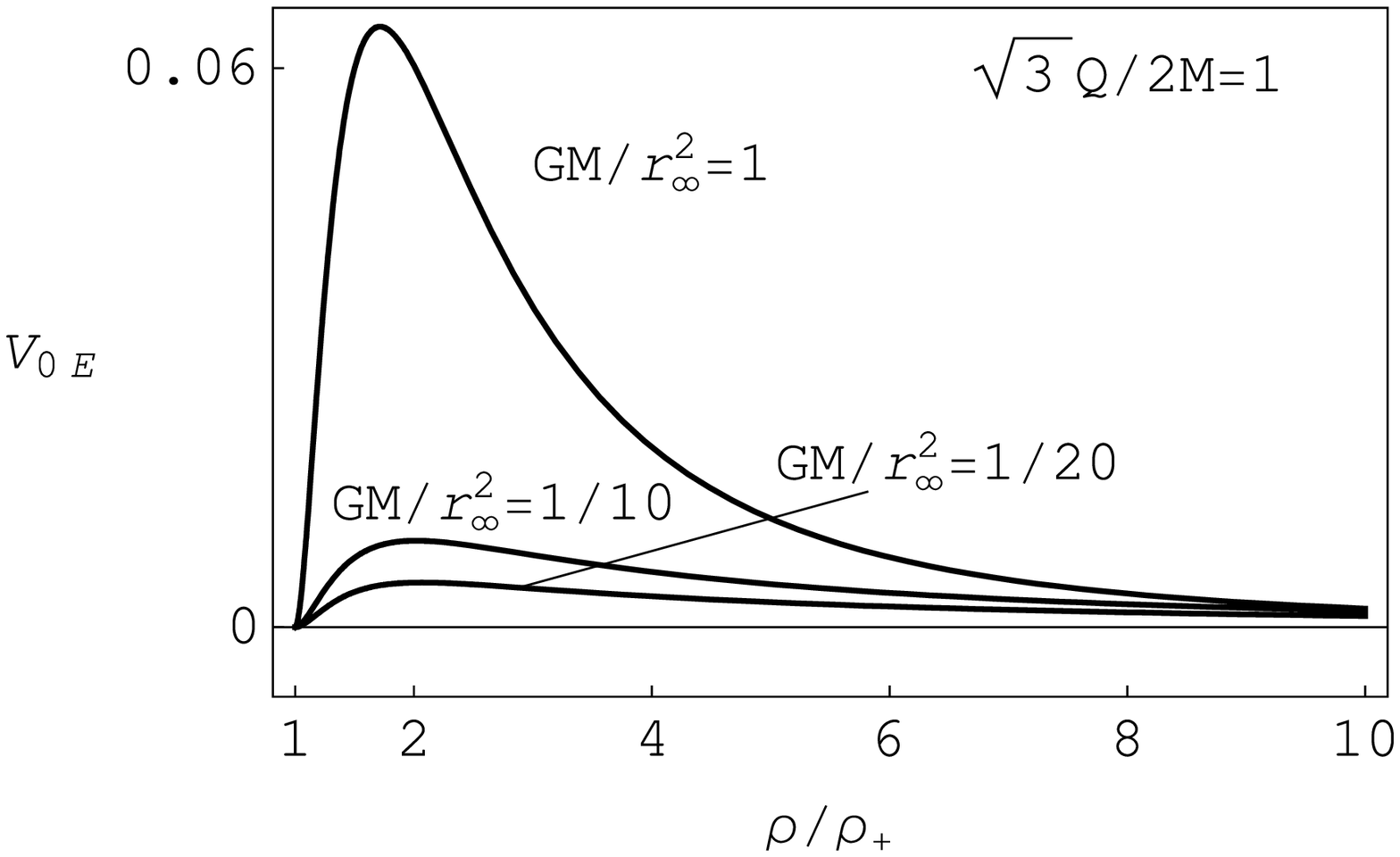}
 \end{center}
 \caption{
The effective potential $V_{0E}$ in maximally charged case.
}
 \label{figK0E;1}
\end{figure}

\begin{figure}[htbp]
 \begin{center}
 \includegraphics[width=9cm,clip]{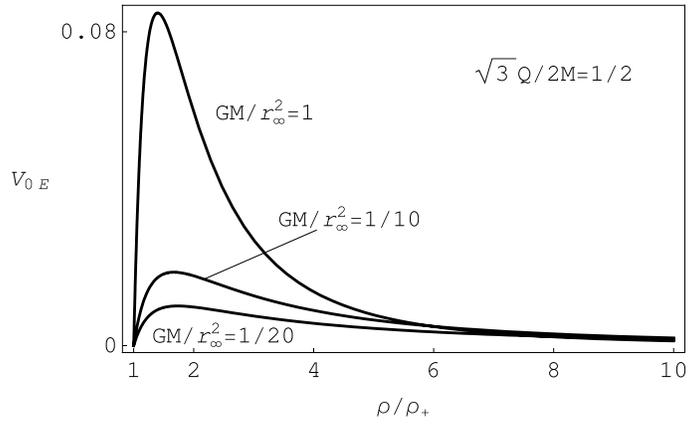}
 \end{center}
 \caption{
The effective potential $V_{0E}$ in $\sqrt{3}Q/2M=1/2$ case.
}
 \label{figK0E;1/2}
\end{figure}

For the $K=0$ mode, we set $h_{\mu \nu}$ and $\delta A_{\mu}$ as
\begin{eqnarray}
h_{\mu \nu}dx^{\mu}dx^{\nu}&=& h_{AB}(\rho)e^{-i\omega t}dx^Adx^B+2h_{A3}(\rho)e^{-i\omega t}dx^A\sigma^3 \notag \\
                           & &+2h_{+-}(\rho)e^{-i\omega t}\sigma^+\sigma^-+h_{33}(\rho)e^{-i\omega t}\sigma^3\sigma^3, \label{eq:K0,met}
\\
\delta A_{\mu}dx^{\mu} &=& \delta A_{A}(\rho)e^{-i\omega t}dx^A+\delta A_3(\rho)e^{-i\omega t}\sigma^3. \label{eq:K0,A}
\end{eqnarray}
We set the gauge $\xi_{\mu}dx^{\mu}$ as
\begin{eqnarray}
\xi_{\mu}dx^{\mu}=\xi_{A}(\rho )e^{-i\omega t}dx^{A}+\xi_{3}(\rho )e^{-i\omega t}\sigma ^3,
\end{eqnarray}
and under the gauge transformation, the metric perturbations transform as
\begin{eqnarray}
h_{tt} &\to & h_{tt}-2i\omega \xi_t-
\frac{(\rho-\rho_-) (\rho-\rho_+) ((\rho_-+\rho_+) \rho-2 \rho_- \rho_+)}{\rho^4(\rho+\rho_0)} \xi_\rho,
\\
h_{t\rho} &\to & h_{t\rho} 
-\frac{(\rho_-+\rho_+) \rho-2 \rho_- \rho_+}{\rho (\rho-\rho_-) (\rho-\rho_+)}\xi_t
-i\omega \xi_{\rho}+\frac{d\xi_t}{d\rho},
\\
h_{\rho \rho} &\to & h_{\rho \rho} + 
\frac{(\rho_-+\rho_++\rho_0) \rho^2+\rho_- \rho_+ (-2 \rho-\rho_0)}
{\rho (\rho-\rho_-) (\rho-\rho_+) (\rho+\rho_0)}
\xi_\rho
+2\frac{d\xi_{\rho}}{d\rho},
\\
h_{t3} &\to & h_{t3}-i\omega \xi_3,
\\
h_{\rho 3} &\to & h_{\rho 3}-\frac{\rho_0}{\rho(\rho +\rho_0)}\xi_3+\frac{d\xi_{3}}{d\rho},
\\
h_{+-} &\to & h_{+-}+\frac{2 (\rho-\rho_-) (\rho-\rho_+) (2 \rho+\rho_0)}{\rho (\rho+\rho_0)}\xi_{\rho},
\\
h_{33} &\to & h_{33}+\frac{(\rho-\rho_-) (\rho-\rho_+) \rho_0 (\rho_-+\rho_0) (\rho_++\rho_0)}{\rho (\rho+\rho_0)^3}\xi_{\rho}.
\end{eqnarray}
So we can choose the condition
\footnote{Note that we can not choose this gauge condition for static perturbation.}
\begin{eqnarray}
h_{+-}=0,\quad h_{tt}=0,\quad h_{t3}=0. \label{eq:K0,gauge}
\end{eqnarray}
In addition, for the gauge transformation (\ref{eq:A,gauge}), the electromagnetic perturbations transform as
\begin{eqnarray}
\delta A_t &\to &\delta A_t-i\omega \chi (\rho),
\\
\delta A_\rho &\to &\delta A_\rho+\frac{d\chi (\rho)}{d\rho},
\\
\delta A_3 &\to &\delta A_3.
\end{eqnarray}
We can choose the gauge condition
\begin{eqnarray}
\delta A_t=0. \label{eq:K0,gauge2}
\end{eqnarray}
Substituting Eqs.  (\ref{eq:K0,met}), (\ref{eq:K0,A}), (\ref{eq:K0,gauge}), and (\ref{eq:K0,gauge2})
into Eqs. (\ref{eq:einstein}) and (\ref{eq:maxwell}),
we obtain perturbation equations whose explicit forms are given in Appendix \ref{appendixA}.
Fortunately, in these modes, the metric and the electromagnetic perturbations are decoupled.
For the metric perturbations, the master equation becomes
\begin{eqnarray}
-\frac{d^2}{d\rho_*^2}\Phi_{0G}+V_{0G}(\rho)\Phi_{0G}=\omega^2\Phi_{0G},
\end{eqnarray}
where the master variable $\Phi_{0G}(\rho)$ and the potential $V_{0G}$ are defined by
\begin{eqnarray}
\Phi_{0G}(\rho):= \frac{(\rho+\rho_0)^{5/4}(2\rho +\rho_0)}{\rho^{1/4}(4\rho+3\rho_0)}h_{33}(\rho),
\end{eqnarray}
\begin{eqnarray}
V_{0G} &=& \frac{(\rho-\rho_+) (\rho-\rho_++\tilde{\rho}_-)}{16 \rho^5 (\tilde{\rho}_0+\tilde{\rho}_--\rho_++\rho)^3 (3(\tilde{\rho}_0+\tilde{\rho}_--\rho_+)+4 \rho)^2}
\notag \\ & & 
\times \bigl[
(128 \tilde{\rho}_0+128 (-\tilde{\rho}_-+3 \rho_+)) 
(\rho-\rho_+)^5
\notag \\ & & 
+(528 \tilde{\rho}_0^2+32 (-\tilde{\rho}_-+55 \rho_+) \tilde{\rho}_0+16(-35 \tilde{\rho}_-^2+98 \rho_+ \tilde{\rho}_-+17 \rho_+^2))(\rho-\rho_+)^4
\notag \\ & & 
+(720 \tilde{\rho}_0^3+48 (13 \tilde{\rho}_-+63 \rho_+)\tilde{\rho}_0^2+16 (-57 \tilde{\rho}_-^2+330 \rho_+ \tilde{\rho}_-+79 \rho_+^2)\tilde{\rho}_0
\notag \\ & & 
+16 (-51 \tilde{\rho}_-^3 
+141 \rho_+ \tilde{\rho}_-^2+63 \rho_+^2 \tilde{\rho}_-+7\rho_+^3)) (\rho-\rho_+)^3
\notag \\ & & 
+(315 \tilde{\rho}_0^4+60(8 \tilde{\rho}_-+41 \rho_+) \tilde{\rho}_0^3+(-450 \tilde{\rho}_-^2 
+6376 \rho_+\tilde{\rho}_-+1850 \rho_+^2) \tilde{\rho}_0^2
\notag \\ & & 
+(-1080 \tilde{\rho}_-^3+5372\rho_+ \tilde{\rho}_-^2+3056 \rho_+^2 \tilde{\rho}_-+460 \rho_+^3) \tilde{\rho}_0
\notag \\ & & 
-465\tilde{\rho}_-^4+35 \rho_+^4+328 \tilde{\rho}_- \rho_+^3+1206 \tilde{\rho}_-^2 \rho_+^2+1456 \tilde{\rho}_-^3\rho_+) (\rho-\rho_+)^2
\notag \\ & & 
+((792 \rho_+-81 \tilde{\rho}_-) \tilde{\rho}_0^4
+12(-27 \tilde{\rho}_-^2+239 \rho_+ \tilde{\rho}_-+98 \rho_+^2)\tilde{\rho}_0^3
\notag \\ & & 
+(-486 \tilde{\rho}_-^3+3852 \rho_+ \tilde{\rho}_-^2+2834 \rho_+^2\tilde{\rho}_-+520 \rho_+^3) \tilde{\rho}_0^2
\notag \\ & & 
+4 (-81 \tilde{\rho}_-^4+567 \rho_+\tilde{\rho}_-^3+535 \rho_+^2 \tilde{\rho}_-^2+209 \rho_+^3 \tilde{\rho}_-+18 \rho_+^4)\tilde{\rho}_0
\notag \\ & & 
+\tilde{\rho}_- (-81 \tilde{\rho}_-^4+492 \rho_+ \tilde{\rho}_-^3
+482 \rho_+^2\tilde{\rho}_-^2+316 \rho_+^3 \tilde{\rho}_-+71 \rho_+^4))(\rho-\rho_+)
\notag \\ & & 
+36 \tilde{\rho}_0^4 \rho_+ (3 \tilde{\rho}_-+8 \rho_+)+12 \tilde{\rho}_0^3 \rho_+ (36\tilde{\rho}_-^2
+ 73 \rho_+ \tilde{\rho}_-+16 \rho_+^2)
\notag \\ & & 
+4 \tilde{\rho}_0^2 \rho_+ (162\tilde{\rho}_-^3+225 \rho_+ \tilde{\rho}_-^2+121 \rho_+^2 \tilde{\rho}_-+8 \rho_+^3)
\notag \\ & & 
+4 \tilde{\rho}_-^2\rho_+ (27 \tilde{\rho}_-^3 
+3 \rho_+ \tilde{\rho}_-^2+25 \rho_+^2 \tilde{\rho}_-+9\rho_+^3)
\notag \\ & & 
+4 \tilde{\rho}_0 \tilde{\rho}_- \rho_+ (108 \tilde{\rho}_-^3+81 \rho_+\tilde{\rho}_-^2+98 \rho_+^2 \tilde{\rho}_-+17 \rho_+^3)
\bigr]. \label{V0G}
\end{eqnarray}
The master equation for the electromagnetic perturbations is given by
\begin{eqnarray}
-\frac{d^2}{d\rho_*^2}\Phi_{0E}+V_{0E}(\rho)\Phi_{0E}=\omega^2\Phi_{0E},
\end{eqnarray}
where $\Phi_{0E}(\rho)$ and $V_{0E}$ are defined by
\begin{eqnarray}
\Phi_{0E}(\rho):= \rho^{1/4}(\rho+\rho_0)^{3/4}\delta A_3(\rho),
\end{eqnarray}
\begin{eqnarray}
V_{0E} &=& \frac{(\rho-\rho_+) (\rho-\rho_++\tilde{\rho}_-)}{16 \rho^5(\tilde{\rho}_0+\tilde{\rho}_--\rho_++\rho)^3}
\bigl[
(8 \tilde{\rho}_0-8 \tilde{\rho}_-+24 \rho_+)
(\rho-\rho_+)^3
\notag \\ & & 
+(15 \tilde{\rho}_0^2+(2 \tilde{\rho}_-+50 \rho_+) \tilde{\rho}_0 
-13\tilde{\rho}_-^2+63 \rho_+^2-2 \tilde{\rho}_- \rho_+)
(\rho-\rho_+)^2
\notag \\ & & 
+( 5(-\tilde{\rho}_-+8 \rho_+) \tilde{\rho}_0^2
+ 10 (-\tilde{\rho}_-^2-3 \rho_+ \tilde{\rho}_- +12\rho_+^2) \tilde{\rho}_0
\notag \\ & & 
+\tilde{\rho}_- (-5 \tilde{\rho}_-^2-70 \rho_+ \tilde{\rho}_-+123\rho_+^2) )
(\rho-\rho_+)
\notag \\ & & 
+4 \tilde{\rho}_-^2 \rho_+ (15 \rho_+ 
-11\tilde{\rho}_-)+4 \tilde{\rho}_0 \tilde{\rho}_- \rho_+ (31 \rho_+-22 \tilde{\rho}_-)+\tilde{\rho}_0^2 (64 \rho_+^2-44\tilde{\rho}_- \rho_+)
\bigr].
\label{V0E}
\end{eqnarray}
From (\ref{ineq:tilde}), (\ref{V0G}) and (\ref{V0E}), we can see $V_{0G}>0$ and $V_{0E}>0$ out side of the horizon. 
The stability has been shown against the perturbation for $K=0$ mode.
Typical profiles of the potentials are plotted in Figs.\ref{figK0G;1}-\ref{figK0E;1/2}.

\subsection{$K=\pm (J+2)$ modes perturbation}
As noted in the previous section, the highest modes for $h_{++}$ and $h_{--}$ are always decoupled for arbitrary $J$. Since these are gauge 
invariant, it is straightforward
to get the perturbation 
equation for $h_{++}$ whose explicit form is given in Appendix \ref{appendixA}.
Defining the new variable
\begin{eqnarray}
\Phi_J(\rho):=  \frac{1}{\rho^{1/4}(\rho+\rho_0)^{3/4}}h_{++}(\rho),
\end{eqnarray}
we obtain the master equation
\begin{eqnarray}
-\frac{d^2}{d\rho_*^2}\Phi_J+V_{J}(\rho)\Phi_J=\omega^2\Phi_J,
\end{eqnarray}
where the potential $V_J(\rho)$ is defined by
\begin{eqnarray}
V_J&=&
\frac{(\rho-\rho_+) (\rho-\rho_++\tilde{\rho}_-)}{16 \tilde{\rho}_0 (\tilde{\rho}_0+\tilde{\rho}_-) \rho^5 (\tilde{\rho}_0+\tilde{\rho}_-+\rho-\rho_+)^3}
\bigl[16 \tilde{\rho}_-^4 (2+J)^2 \rho_+^2
\notag \\ & & 
+4 \tilde{\rho}_-^3 \rho_+ (3 \tilde{\rho}_- 
+(57+60 J+16 J^2) \rho_+) \tilde{\rho}_0+4 \tilde{\rho}_-^2 \rho_+ (9 \tilde{\rho}_-+2(37+42 J+12 J^2) \rho_+) \tilde{\rho}_0^2
\notag \\ & & 
+4 \tilde{\rho}_- \rho_+  
(9 \tilde{\rho}_-+(41+52 J+16 J^2) \rho_+) \tilde{\rho}_0^3+4 \rho_+ (3 \tilde{\rho}_-+4(2+3 J+J^2) \rho_+) \tilde{\rho}_0^4 
\notag \\& &
+(32 \tilde{\rho}_-^3 (2+J)^2 \rho_+ (\tilde{\rho}_-+2 \rho_+)+\tilde{\rho}_-^2 (-9 \tilde{\rho}_-^2+2 \tilde{\rho}_- (249+240 J+64 J^2)\rho_+
\notag \\ & & 
+(711 
+736 J+192 J^2) \rho_+^2) \tilde{\rho}_0+\tilde{\rho}_- (-27 \tilde{\rho}_-^2+4 \tilde{\rho}_- (175+168 J+48 J^2)\rho_+
\notag \\ & & 
+(655+704 J 
+192 J^2) \rho_+^2) \tilde{\rho}_0^2+(-27 \tilde{\rho}_-^2+2 \tilde{\rho}_- (209+208 J+64 J^2) \rho_+
\notag \\ & & 
+8(25+28 J 
+8 J^2) \rho_+^2) \tilde{\rho}_0^3+(-9 \tilde{\rho}_-+8 (11+12 J+4 J^2) \rho_+) \tilde{\rho}_0^4) (\rho-\rho_+) 
\notag \\ & & 
+(16 \tilde{\rho}_-^2 (2+J)^2 (\tilde{\rho}_-^2 
+8 \tilde{\rho}_- \rho_++6 \rho_+^2)+\tilde{\rho}_- (\tilde{\rho}_-^2 (207+240 J+64 J^2)
\notag \\ & & 
+2 \tilde{\rho}_-(735+736 J+192 J^2) \rho_++(739 
+752 J+192 J^2) \rho_+^2) \tilde{\rho}_0
\notag \\ & & 
+(\tilde{\rho}_-^2 (257+336 J+96J^2)+8 \tilde{\rho}_- (175+176 J+48 J^2) \rho_+ 
\notag \\ & & 
+(355+368 J+96 J^2) \rho_+^2) \tilde{\rho}_0^2+(\tilde{\rho}_- (149+208J+64 J^2)
\notag \\ & & 
+2 (221+224 J+64 J^2)  
\rho_+) \tilde{\rho}_0^3+(35+48 J+16 J^2) \tilde{\rho}_0^4)(\rho-\rho_+)^2
\notag \\ & & 
+(64 \tilde{\rho}_- (2+J)^2 (\tilde{\rho}_-^2+3 \tilde{\rho}_- \rho_++\rho_+^2) 
+8 (\tilde{\rho}_-^2 (87+92 J+24 J^2)
\notag \\ & & 
+\tilde{\rho}_- (187+188J+48 J^2) \rho_++8 (2+J)^2 \rho_+^2) \tilde{\rho}_0+8 (8 \tilde{\rho}_- 
(10+11 J+3 J^2)
\notag \\ & & 
+(91+92 J+24 J^2)\rho_+) \tilde{\rho}_0^2+8 (25+28 J+8 J^2) \tilde{\rho}_0^3) (\rho-\rho_+)^3 
\notag \\& &
+(16 (2+J)^2 (6 \tilde{\rho}_-^2+8 \tilde{\rho}_- \rho_++\rho_+^2)+16 (2+J) (\tilde{\rho}_- (23+12 J)+8 (2+J) \rho_+) \tilde{\rho}_0 
\notag \\& &
+16(22+23 J+6 J^2) \tilde{\rho}_0^2) (\rho-\rho_+)^4
\notag \\ & & 
+(32 (2+J)^2 (2 \tilde{\rho}_-+\rho_+)+64 (2+J)^2\tilde{\rho}_0)  
(\rho-\rho_+)^5
+16 (2+J)^2 (\rho-\rho_+)^6\bigr]. 
\label{V_J}
\end{eqnarray}
Since we can see $V_J>0$ out side of the horizon from (\ref{ineq:tilde}) and (\ref{V_J}),
the stability has been shown against perturbation for $K=\pm (J+2)$ modes.

\section{SUMMARY AND DISCUSSION}
In this paper, we have extended the stability analysis for perturbations
of the SqKK black holes given in~\cite{Kimura:2007cr} to
the SqKK black holes with charge~\cite{Ishihara:2005dp}.
We have expanded the perturbation variables in terms of
Wigner function $D^{J}_{KM}$ which have three indices $J,M,K$,
and derived the master equations for $(J=0,M=0,K=0,\pm 1,\pm 2)$ modes which have $SU(2)$ symmetry
and the highest modes $(K=\pm (J+2) )$.
Using the effective potential functions in the master equations,
we have discussed the stability for these modes.

Since the modes of $J=0,M=0,K=\pm 2$ and the highest modes contain only the gravitational perturbation,
the perturbation equation for each mode can be reduced to a single master equation.
In the mode of $J=0,M=0,K=0$,
the perturbed equations 
can be reduced to two decoupled master equations.
As for the modes of $J=0,M=0,K=\pm 1$, we have obtained the coupled master equations 
for the gravitational and the electromagnetic perturbations, in contrast to the case for
Reissner-Nordstr\"{o}m black holes of which these perturbations are decoupled against all modes~\cite{Kodama:2003kk}.
We guess that the coupling for the gravitational and the electromagnetic perturbations comes from the differences of the asymptotic structures.

We have shown from the positivity of the effective potential functions of the master equations
that the modes of $J=0,M=0,K=0,\pm 2$ and $K=\pm(J+2)$ are stable as those in the neutral case.
In the case of $J=0,M=0,K=\pm 1$, the effective potential functions form a matrix.
In maximally charged case,
we have shown the stability for $J = 0, K=\pm 1$ by using the matrix of effective potential functions.
In general, because there are domains in which the effective potential functions are negative, 
we must find an appropriate S-deformation function
similar to the discussions in~\cite{Ishibashi:2003ap,Kodama:2003kk}
to show the stability for $J=0,M=0,K=\pm 1$ modes.
Even in neutral cases where the gravitational and the electromagnetic perturbations are decoupled,
the S-deformation function have not been found yet.
It is hard to show the stability of these modes in the case of charged black hole.
To show stability for $J = 0, K=\pm 1$ modes in general is open issue.

Since the stability for $J=0$ modes is suggested in the limit of both neutral
case~\cite{Kimura:2007cr,Ishihara:2008re}
and extremal case,
and the instability empirically appears in the lower modes,
we expect SqKK black holes with charge are stable.

\section*{Acknowledgments}
We would like to thank
Akihiro Ishibashi, Hideki Ishihara, Hideo Kodama, Keiju Murata, Ken-ichi Nakao and Jiro Soda
for useful discussions.
M.K. is partially supported by the JSPS Grant-in-Aid for Scientific Research No. 20$\cdot$7858.

\appendix
\allowdisplaybreaks
\section{the perturbation equations}\label{appendixA}
In this section, the precise form of the perturbation
equations are given for each mode.
\subsection{$K=\pm 2$ modes}
In $K=2$ mode, there is only $(++)$ component of linearized Einstein equation 
which is given by
\begin{eqnarray}
&& \delta G_{++}-2\delta T_{++} \notag \\
&=&\frac{h_{++}}{2\rho ^3(\rho+\rho_0)^3(\rho_0+\rho_+)(\rho_0+\rho_-)}[4\rho^6+16\rho^5\rho_0-2\rho_0^2\rho_+\rho_-(\rho_0+\rho_+)(\rho_0+\rho_-) \notag \\
& &+4\rho^4(5\rho_0^2-\rho_+\rho_--\rho_0(\rho_++\rho_-))+\rho \rho_0(\rho_0+\rho_+)(\rho_0+\rho_-)(-6\rho_+\rho_-+\rho_0(\rho_++\rho_-)) \notag \\
& &+4\rho^3(3\rho_0^3+\rho_+\rho_-(\rho_++\rho_-)+\rho_0(\rho_+^2+\rho_+\rho_-+\rho_-^2))+\rho^2(2\rho_0^4-6\rho_+^2\rho_-^2+\rho_0^3(\rho_++\rho_-) \notag \\
& &-3\rho_0\rho_+\rho_-(\rho_++\rho_-)+\rho_0^2(3\rho_+^2+3\rho_+^2-2\rho_+\rho_-))] \notag \\
& &+\frac{2\rho^3+2\rho_0\rho_+\rho_--3\rho^2(\rho_++\rho_-)-\rho(-4\rho_+\rho_-+\rho_0(\rho_++\rho_-))}{2\rho^2(\rho+\rho_0)^2}\frac{dh_{++}}{d\rho} \notag \\
& &-\frac{(\rho-\rho_+)(\rho-\rho_-)}{2\rho(\rho+\rho_0)}\frac{d^2h_{++}}{d\rho^2}-\frac{\rho^2}{2(\rho-\rho_+)(\rho-\rho_-)}\omega^2h_{++}=0.
\label{Eeq_pp}
\end{eqnarray}
The equation for $K = -2$ mode can be obtained
by just replacing $h_{++}$ to $h_{--}$ in Eq. (\ref{Eeq_pp}).

\subsection{$K=\pm 1$ modes}
In $K= 1$ mode, 
there are $(t+), (\rho +)$ and $(+3)$ components of linearized Einstein equation 
and there is $(+)$ component of Maxwell equation for perturbations
which are given by 
\begin{eqnarray}
& &\delta G_{t+}-2 \delta T_{t+} \notag \\
&=& \frac{1}{2 \rho^3 (\rho_-+\rho_0) (\rho_++\rho_0) (\rho+\rho_0)^2}[\rho^5+3 \rho_0 \rho^4+3 \rho_0^2 \rho^3+(\rho_0^3-2 (\rho_-+\rho_+) \rho_0^2-2 (\rho_-+ \notag \\
& & \rho_+)^2 \rho_0-2 \rho_- \rho_+ (\rho_-+\rho_+)) \rho^2-(\rho_-+\rho_0)(\rho_++\rho_0) ((\rho_-+\rho_+) \rho_0 \notag \\
& & -3 \rho_- \rho_+) \rho+\rho_- \rho_+ \rho_0 (\rho_-+\rho_0) (\rho_++\rho_0)]h_{t+}-\frac{(\rho-\rho_-) (\rho-\rho_+) \rho_0}{2 \rho^2 (\rho+\rho_0)^2}\frac{dh_{t+}}{d\rho} \notag \\
& & -\frac{(\rho-\rho_-) (\rho-\rho_+)}{2 \rho (\rho+\rho_0)}\frac{d^2h_{t+}}{d\rho^2}-\frac{i\omega (\rho-\rho_-) (\rho-\rho_+)}{\rho^2 (\rho+\rho_0)}h_{\rho+}-\frac{i\omega (\rho-\rho_-) (\rho-\rho_+)}{2 \rho (\rho+\rho_0)}\frac{dh_{\rho+}}{d\rho} \notag \\
& & +\frac{\sqrt{3} \sqrt{\rho_- \rho_+} (\rho_--\rho) (\rho_+-\rho)}{\rho^3(\rho+\rho_0)}\frac{d\delta A_+}{d\rho}=0,
\label{Eeq_tp}
\\ & & \notag \\ 
& &\delta G_{\rho +}-2 \delta T_{\rho +} \notag \\
&=& \frac{(\rho^2+2 \rho_0 \rho-\rho_- \rho_+-(\rho_-+\rho_+) \rho_0)^2}{2 \rho (\rho_-+\rho_0) (\rho_++\rho_0) (\rho+\rho_0)^3}h_{\rho+}-\frac{i\omega \rho (2 \rho+\rho_0)}{2 (\rho-\rho_-) (\rho-\rho_+) (\rho+\rho_0)}h_{t+} \notag \\
& & -\frac{i\omega \sqrt{3} \sqrt{\rho_- \rho_+}}{(\rho-\rho_-) (\rho-\rho_+)}\delta A_++\frac{i\omega \rho^2}{2 (\rho-\rho_-) (\rho-\rho_+)}\frac{dh_{t+}}{d\rho}-\frac{\omega ^2\rho^2}{2 (\rho-\rho_-) (\rho-\rho_+)}h_{\rho+}=0, 
\label{Eeq_rp}
\\ & & \notag \\ 
& &\delta G_{+3}-2 \delta T_{+3} \notag \\
&=& -\frac{i}{2 \rho^2 (\rho+\rho_0)^4}[(2 \rho^5+(-\rho_--\rho_++5 \rho_0) \rho^4+2 (3\rho_0^2+\rho_- \rho_+) \rho^3-(5 (\rho_-+\rho_+)\rho_0^2 \notag \\
& & +(3 \rho_-^2+8 \rho_+ \rho_-+3 \rho_+^2) \rho_0+3 \rho_- \rho_+(\rho_-+\rho_+)) \rho^2+2 \rho_- \rho_+ (\rho_0^2+2 (\rho_-+\rho_+)\rho_0+2 \rho_- \rho_+) \rho \notag \\
& & +\rho_- \rho_+ \rho_0 (\rho_- \rho_++(\rho_-+\rho_+)\rho_0))]h_{\rho+} +\frac{\omega \rho^2 (\rho^2+2 \rho_0 \rho-\rho_- \rho_+-(\rho_-+\rho_+)\rho_0)}{2 (\rho-\rho_-) (\rho-\rho_+) (\rho+\rho_0)^2}h_{t+}\notag \\
& & -\frac{i (\rho-\rho_-) (\rho-\rho_+) (\rho^2+2 \rho_0 \rho-\rho_- \rho_+-(\rho_-+\rho_+) \rho_0)}{2 \rho (\rho+\rho_0)^3}\frac{dh_{\rho+}}{d\rho}=0,
\label{Eeq_p3}
\\ & & \notag \\ 
& &\delta(\nabla _{\mu}F^{\mu +}) \notag \\
&=& \frac{-1}{\sin^2\theta}[\frac{\delta A_+}{\rho^2 (\rho_-+\rho_0) (\rho_++\rho_0)}-\frac{((\rho_-+\rho_++\rho_0) \rho^2-2 \rho_- \rho_+ \rho-\rho_- \rho_+ \rho_0)}{\rho^3 (\rho+\rho_0)^3}\frac{d\delta A_+}{d\rho} \notag \\
& & -\frac{\omega ^2\rho}{(\rho-\rho_-) (\rho-\rho_+) (\rho+\rho_0)}\delta A_+-\frac{\sqrt{3} \sqrt{\rho_- \rho_+} (\rho_--\rho) (\rho_+-\rho) (2\rho+\rho_0)}{2 \rho^3 (\rho-\rho_-) (\rho-\rho_+) (\rho+\rho_0)^3}h_{t+} \notag \\
& & +\frac{\sqrt{3} \sqrt{\rho_- \rho_+} (\rho_--\rho) (\rho_+-\rho)}{2 \rho^2(\rho-\rho_-) (\rho-\rho_+) (\rho+\rho_0)^2}\frac{dh_{t+}}{d\rho}+\frac{i\omega \sqrt{3} \sqrt{\rho_- \rho_+} (\rho_--\rho) (\rho_+-\rho)}{2 \rho^2(\rho-\rho_-) (\rho-\rho_+) (\rho+\rho_0)^2}h_{\rho+} \notag \\
& & -\frac{(\rho-\rho_-) (\rho-\rho_+)}{\rho^2 (\rho+\rho_0)^2}\frac{d^2\delta A_+}{d\rho^2}]=0.
\label{Eeq_p}
\end{eqnarray}
The equations for $K = -1$ mode can be obtained by taking 
Eq. (\ref{Eeq_tp}), (\ref{Eeq_rp}), (\ref{Eeq_p3}), and (\ref{Eeq_p})
to its complex conjugate equations, and using the relations $\bar{h}_{A+}=h_{A-}$, $\bar{h}_{+3}=h_{-3}$ and $\delta \bar{A}_+=\delta A_-$.

\subsection{$K=0$ mode}
In $K=0$ mode, 
there are $(tt), (t\rho ), (\rho \rho ), (t3), (\rho 3), (+-)$ and $(33)$ components of linearized Einstein equation 
and there are $(t), (\rho )$ and $(3)$ components of Maxwell equation for perturbations
which are given by 
\begin{eqnarray}
& &\delta G_{tt}-2 \delta T_{tt} \notag \\
&=&-\frac{\rho_- \rho_+ (\rho-\rho_-) (\rho-\rho_+)}{4 \rho^6 (\rho_-+\rho_0) (\rho_++\rho_0)}h_{33}+\frac{(\rho-\rho_-)^2 (\rho-\rho_+)^2}{4 \rho^6 (\rho+\rho_0)^4}[4 \rho^4+4 (\rho_-+\rho_++3 \rho_0) \rho^3 \notag \\
& &+(6 \rho_0^2+2 (\rho_-+\rho_+) \rho_0-9 \rho_- \rho_+) \rho^2-10 \rho_- \rho_+ \rho_0 \rho-3 \rho_- \rho_+ \rho_0^2]h_{\rho\rho} \notag \\
& &-\frac{(\rho-\rho_-) (\rho-\rho_+) (4 \rho^3+(\rho_0-3 (\rho_-+\rho_+)) \rho^2+2 \rho_- \rho_+ \rho-\rho_- \rho_+ \rho_0)}{4 \rho^5 (\rho_-+\rho_0) (\rho_++\rho_0) (\rho+\rho_0)}\frac{dh_{33}}{d\rho} \notag \\
& &+\frac{(\rho-\rho_-)^3 (\rho-\rho_+)^3 (4 \rho+3 \rho_0)}{4 \rho^5 (\rho+\rho_0)^3}\frac{dh_{\rho\rho}}{d\rho}-\frac{(\rho-\rho_-)^2 (\rho-\rho_+)^2}{2 \rho^4 (\rho_-+\rho_0) (\rho_++\rho_0)}\frac{d^2h_{33}}{d\rho^2} \notag \\
& &+\frac{i\omega \sqrt{3} \sqrt{\rho_- \rho_+} (\rho_--\rho) (\rho_+-\rho)}{\rho^3 (\rho+\rho_0)}\delta A_{\rho}=0,
\\ && \notag\\
& & \delta G_{t\rho}-2 \delta T_{t\rho} \notag \\
&=& \frac{-i\omega (\rho-\rho_+) (\rho-\rho_-) (3 \rho_0+4 \rho)}{4 \rho^2 (\rho_0+\rho)^2}h_{\rho\rho}+\frac{i\omega (\rho_0 (\rho_+ \rho_--\rho^2)-\rho (\rho_- \rho+\rho_+(\rho-2 \rho_-)))}{4 (\rho_0+\rho_+) (\rho_0+\rho_-) (\rho_+-\rho)(\rho_--\rho) \rho^2}h_{33}\notag \\
& & +\frac{i\omega (\rho_0+\rho)}{2 (\rho_0+\rho_+) (\rho_0+\rho_-) \rho}\frac{dh_{33}}{d\rho}=0,
\\ && \notag\\
& &\delta G_{\rho \rho}-2 \delta T_{\rho \rho} \notag \\
&=& \frac{(\rho_--\rho) \rho_0 (4 \rho+\rho_0)+\rho_+ (\rho_0 (\rho_0-2 \rho_-)+\rho (\rho_-+4 \rho_0))}{4 \rho^2 (\rho-\rho_-) (\rho-\rho_+) (\rho_-+\rho_0) (\rho_++\rho_0) (\rho+\rho_0)}h_{33} \notag \\
& & +\frac{-4 \rho^2-8 \rho_0 \rho-3 \rho_0^2+\rho_- \rho_++(\rho_-+\rho_+) \rho_0}{4 \rho (\rho+\rho_0)^3}h_{\rho\rho} \notag \\
& & +\frac{4 \rho^2+(2 \rho_0-3 (\rho_-+\rho_+)) \rho+2 \rho_- \rho_+-(\rho_-+\rho_+) \rho_0}{4 \rho (\rho-\rho_-) (\rho-\rho_+) (\rho_-+\rho_0) (\rho_++\rho_0)}\frac{dh_{33}}{d\rho} \notag \\
& & -\frac{i\omega \rho (4 \rho+3 \rho_0)}{2 (\rho-\rho_-) (\rho-\rho_+) (\rho+\rho_0)}h_{t\rho}+\frac{\omega ^2 \rho^2 (\rho+\rho_0)^2}{2 (\rho-\rho_-)^2 (\rho-\rho_+)^2 (\rho_-+\rho_0) (\rho_++\rho_0)}h_{33} \notag \\
& & -\frac{i \omega \sqrt{3} \sqrt{\rho_- \rho_+}}{(\rho-\rho_-) (\rho-\rho_+)}\delta A_{\rho}=0, 
\\ && \notag\\
& &\delta G_{t3}-2 \delta T_{t3}\notag \\
&=&\frac{\sqrt{3} \sqrt{\rho_- \rho_+} (\rho_--\rho) (\rho_+-\rho)}{\rho^3 (\rho+\rho_0)}\frac{d\delta A_3}{d\rho}-\frac{i\omega (\rho-\rho_-) (\rho-\rho_+)}{\rho^2 (\rho+\rho_0)}h_{\rho3} \notag \\
& &-\frac{i\omega (\rho-\rho_-) (\rho-\rho_+)}{2 \rho (\rho+\rho_0)}\frac{dh_{\rho3}}{d\rho}=0, 
\\ && \notag\\
& &\delta G_{\rho 3}-2 \delta T_{\rho 3}\notag \\
&=&-\frac{i\omega \sqrt{3} \sqrt{\rho_- \rho_+}}{(\rho-\rho_-) (\rho-\rho_+)}\delta A_3-\frac{\omega ^2\rho^2}{2 (\rho-\rho_-) (\rho-\rho_+)}h_{\rho3}=0,
\\ && \notag\\
& &\delta G_{+-}-2 \delta T_{+-} \notag \\
&=&\frac{2 \rho_0 \rho^3-(\rho_0^2+5 (\rho_-+\rho_+) \rho_0+\rho_- \rho_+) \rho^2-2 \rho_0 ((\rho_-+\rho_+) \rho_0-3 \rho_- \rho_+) \rho+4 \rho_- \rho_+ \rho_0^2}{2 \rho^3 (\rho_-+\rho_0) (\rho_++\rho_0) (\rho+\rho_0)}h_{33} \notag \\
& &+\frac{1}{2 \rho^2 (\rho+\rho_0)^3}[-2 (\rho_-+\rho_++\rho_0) \rho^4+(\rho_-^2+5 \rho_+ \rho_-+\rho_+^2-3 \rho_0^2-2 (\rho_-+\rho_+) \rho_0) \rho^3 \notag \\
& &+(2 (\rho_-+\rho_+) \rho_0^2+2 (\rho_-^2+5 \rho_+ \rho_-+\rho_+^2) \rho_0-\rho_- \rho_+ (\rho_-+\rho_+)) \rho^2 \notag \\
& &+\rho_- \rho_+ (\rho_0^2-4 (\rho_-+\rho_+) \rho_0-\rho_- \rho_+) \rho -\rho_- \rho_+ (\rho_-+\rho_+) \rho_0^2]h_{\rho\rho} \notag \\
& &+\frac{\rho^3+((\rho_-+\rho_+) \rho_0-\rho_- \rho_+) \rho-2 \rho_- \rho_+ \rho_0}{\rho^2 (\rho_-+\rho_0) (\rho_++\rho_0)}\frac{dh_{33}}{d\rho}+2i\omega \sqrt{3} \sqrt{\rho_- \rho_+}\delta A_{\rho} \notag \\
& &-\frac{(\rho-\rho_-) (\rho-\rho_+) (-\rho_--\rho_++2 \rho)}{2 \rho (\rho+\rho_0)}\frac{dh_{\rho\rho}}{d\rho}+\frac{(\rho-\rho_-) (\rho-\rho_+) (\rho+\rho_0)}{\rho (\rho_-+\rho_0) (\rho_++\rho_0)}\frac{d^2h_{33}}{d\rho^2} \notag \\
& &+\frac{\omega ^2 \rho^2 (\rho+\rho_0)^2}{(\rho-\rho_-) (\rho-\rho_+) (\rho_-+\rho_0) (\rho_++\rho_0)}h_{33}+\omega ^2\rho ^2 h_{\rho \rho}-2i\omega \rho^2\frac{dh_{t\rho}}{d\rho} \notag \\
& &-\frac{i\omega \rho (2 \rho^3+(-\rho_--\rho_++3 \rho_0) \rho^2-2 (\rho_-+\rho_+) \rho_0 \rho+\rho_- \rho_+ \rho_0)}{(\rho-\rho_-) (\rho-\rho_+) (\rho+\rho_0)}h_{t\rho}=0, 
\\ && \notag\\
& &\delta G_{33}-2 \delta T_{33} \notag \\
&=& -\frac{(\rho_-+\rho_0) (\rho_++\rho_0)}{4 \rho^2 (\rho+\rho_0)^5} [4 \rho^5+12 \rho_0 \rho^4+(-3 \rho_-^2-13 \rho_+ \rho_--3 \rho_+^2+3 \rho_0^2-12 (\rho_-+\rho_+) \rho_0) \rho^3 \notag \\
& & +(-2 (\rho_-+\rho_+) \rho_0^2+2(\rho_-^2+3 \rho_+ \rho_-+\rho_+^2) \rho_0+11 \rho_- \rho_+ (\rho_-+\rho_+)) \rho^2-\rho_- \rho_+ (\rho_0^2 \notag \\
& & -2 (\rho_-+\rho_+) \rho_0+7 \rho_- \rho_+) \rho+\rho_- \rho_+ \rho_0 ((\rho_-+\rho_+) \rho_0-2 \rho_- \rho_+)]h_{\rho\rho} \notag \\
& & +\frac{(\rho-\rho_-) (\rho-\rho_+) (\rho_-+\rho_0) (\rho_++\rho_0)}{4 \rho (\rho+\rho_0)^4}[(-4 \rho^2+(3 (\rho_-+\rho_+)-2 \rho_0) \rho \notag \\
& & -2 \rho_- \rho_++(\rho_-+\rho_+) \rho_0)]\frac{dh_{\rho\rho}}{d\rho}+\frac{3 (\rho_-+\rho_0) (\rho_++\rho_0)}{4 \rho (\rho+\rho_0)^3}h_{33} \notag \\
& & -\frac{i\omega \rho (\rho_-+\rho_0) (\rho_++\rho_0)}{2 (\rho-\rho_-) (\rho-\rho_+) (\rho+\rho_0)^3}[(4 \rho^3+3 (-\rho_--\rho_++\rho_0) \rho^2-2 ((\rho_-+\rho_+) \rho_0-\rho_- \rho_+) \rho \notag \\
& & +\rho_- \rho_+ \rho_0)]h_{t\rho}+\frac{i\omega \sqrt{3} \sqrt{\rho_- \rho_+} (\rho_-+\rho_0) (\rho_++\rho_0)}{(\rho+\rho_0)^2}\delta A_{\rho} \notag \\
& & -\frac{i\omega \rho^2 (\rho_-+\rho_0) (\rho_++\rho_0)}{(\rho+\rho_0)^2}\frac{dh_{t\rho}}{d\rho}+\frac{\omega ^2\rho^2 (\rho_-+\rho_0) (\rho_++\rho_0)}{2 (\rho+\rho_0)^2}h_{\rho\rho}=0, \notag \\
& &
\\ & & \notag \\ 
& &\delta(\nabla _{\mu}F^{\mu t}) \notag \\
&=&\frac{-1}{4\rho^3(\rho +\rho_0)^3(\rho_0 +\rho_+)(\rho_0 +\rho_-)} \notag \\
& &[\sqrt{3} \sqrt{\rho_- \rho_+} \rho_0 (\rho+\rho_0)^2 h_{33}+\sqrt{3} \sqrt{\rho_- \rho_+} (\rho_-+\rho_0) (\rho_++\rho_0) ((\rho_-+\rho_++\rho_0) \rho^2 \notag \\
& &-2 \rho_- \rho_+ \rho-\rho_- \rho_+ \rho_0)h_{\rho\rho}-\sqrt{3} \sqrt{\rho_- \rho_+} \rho (\rho+\rho_0)^3\frac{dh_{33}}{d\rho}+\sqrt{3} \sqrt{\rho_- \rho_+} (\rho_--\rho) (\rho_+-\rho) \rho \notag \\
& & (\rho_-+\rho_0) (\rho_++\rho_0) (\rho+\rho_0)\frac{dh_{\rho\rho}}{d\rho}+i\omega 8 \rho^3 (\rho_-+\rho_0) (\rho_++\rho_0) (\rho+\rho_0)^2\delta A_{\rho} \notag \\
& & +i\omega 4 \rho^4 (\rho_-+\rho_0) (\rho_++\rho_0) (\rho+\rho_0)^2\frac{d\delta A_\rho}{d\rho}]=0,
\\ && \notag\\
& &\delta(\nabla _{\mu}F^{\mu \rho}) \notag \\
&=&\frac{-1}{4\rho^2(\rho +\rho_0)^2(\rho_0 +\rho_+)(\rho_0 +\rho_-)} \notag \\
& &[-i\omega \sqrt{3} \sqrt{\rho_- \rho_+} (\rho+\rho_0)^2h_{33}+i\omega \sqrt{3} \sqrt{\rho_- \rho_+} (\rho_--\rho) (\rho_+-\rho) (\rho_-+\rho_0) (\rho_++\rho_0)h_{\rho\rho} \notag \\
& &-\omega ^24 \rho^3 (\rho_-+\rho_0) (\rho_++\rho_0) (\rho+\rho_0)\delta A_\rho ]=0,
\\ && \notag\\
& &\delta(\nabla _{\mu}F^{\mu 3}) \notag \\
&=& -\frac{i\omega \sqrt{3} \sqrt{\rho_- \rho_+} }{2 \rho^2  (\rho_-+\rho_0) (\rho_++\rho_0)}h_{\rho 3}-\frac{1}{\rho^2 (\rho+\rho_0)^2}\delta A_3 \notag \\
& & +\frac{2 \rho^3+(-\rho_--\rho_++\rho_0) \rho^2-\rho_- \rho_+ \rho_0}{\rho^3 (\rho_-+\rho_0) (\rho_++\rho_0) (\rho+\rho_0)}\frac{d\delta A_3}{d\rho}+\frac{(\rho-\rho_-) (\rho-\rho_+)}{\rho^2 (\rho_-+\rho_0) (\rho_++\rho_0)}\frac{d^2\delta A_3}{d\rho^2}\notag \\
& & +\frac{\omega ^2 \rho (\rho+\rho_0)}{(\rho-\rho_-) (\rho-\rho_+) (\rho_-+\rho_0) (\rho_++\rho_0)}\delta A_3=0.
\end{eqnarray}

\subsection{$K= \pm (J + 2)$ modes}
In $K=(J+2)$ mode, there is only $(++)$ component of linearized Einstein equation 
which is given by 
\begin{eqnarray}
& &\delta G_{++}-2 \delta T_{++} \notag \\
 &=& \frac{h_{++}}{2 \rho^3 (\rho_-+\rho_0) (\rho_++\rho_0) (\rho+\rho_0)^3}[(J+2)^2 \rho^6+4 (J+2)^2 \rho_0 \rho^5+\rho^4 ((2 J+5) (3 J+4) \rho_0^2 \notag \\
 & & -(J+4) (\rho_-+\rho_+)\rho_0-(J+4) \rho_- \rho_+)+2 \rho^3 ((J+2) (2 J+3) \rho_0^3-J (\rho_-+\rho_+)\rho_0^2 \notag \\
 & & +(2 \rho_-^2-(J-2) \rho_+ \rho_-+2 \rho_+^2) \rho_0+2\rho_- \rho_+ (\rho_-+\rho_+))+\rho^2 ((J+1) (J+2) \rho_0^4 \notag \\              
 & & -(J-1) (\rho_-+\rho_+)\rho_0^3+(3 \rho_-^2-(J+2) \rho_+ \rho_-+3 \rho_+^2)\rho_0^2-3 \rho_- \rho_+ (\rho_-+\rho_+) \rho_0-6 \rho_-^2 \rho_+^2) \notag \\
 & & +\rho \rho_0 (\rho_-+\rho_0) (\rho_++\rho_0) ((\rho_-+\rho_+) \rho_0-6 \rho_- \rho_+)-2 \rho_- \rho_+ \rho_0^2 (\rho_-+\rho_0) (\rho_++\rho_0)] \notag \\
 & & +\frac{2 \rho^3-3 (\rho_-+\rho_+) \rho^2-((\rho_-+\rho_+) \rho_0-4 \rho_- \rho_+) \rho+2 \rho_- \rho_+ \rho_0}{2 \rho^2 (\rho+\rho_0)^2}\frac{dh_{++}}{d\rho} \notag \\
 & & -\frac{(\rho-\rho_-) (\rho-\rho_+)}{2 \rho (\rho+\rho_0)}\frac{d^2h_{++}}{d\rho ^2}-\frac{\rho^2}{2 (\rho-\rho_-) (\rho-\rho_+)}\omega ^2h_{++}=0.
\label{J_pp}
\end{eqnarray}
The equation for $K = -(J + 2)$ mode can be obtained
by replacing $h_{++}$ to $h_{--}$ in Eq. (\ref{J_pp}).



\begin{thebibliography}{99}

\bibitem{Banks:1999gd}
  T.~Banks and W.~Fischler,
  arXiv:hep-th/9906038.

\bibitem{Dimopoulos:2001hw}
  S.~Dimopoulos and G.~L.~Landsberg,
  Phys.\ Rev.\ Lett.\  {\bf 87}, 161602 (2001)
  [arXiv:hep-ph/0106295].

\bibitem{Giddings:2001bu}
  S.~B.~Giddings and S.~D.~Thomas,
  Phys.\ Rev.\  D {\bf 65}, 056010 (2002)
  [arXiv:hep-ph/0106219].

\bibitem{Ida:2002ez}
  D.~Ida, K.~y.~Oda and S.~C.~Park,
  Phys.\ Rev.\  D {\bf 67}, 064025 (2003)
  [Erratum-ibid.\  D {\bf 69}, 049901 (2004)]
  [arXiv:hep-th/0212108].

\bibitem{Ida:2005ax}
  D.~Ida, K.~y.~Oda and S.~C.~Park,
  Phys.\ Rev.\  D {\bf 71}, 124039 (2005)
  [arXiv:hep-th/0503052].

\bibitem{Ida:2006tf}
  D.~Ida, K.~y.~Oda and S.~C.~Park,
  Phys.\ Rev.\  D {\bf 73}, 124022 (2006)
  [arXiv:hep-th/0602188].

\bibitem{Myers:1986un}
  R.~C.~Myers and M.~J.~Perry,
  Annals Phys.\  {\bf 172}, 304 (1986).

\bibitem{Emparan:2001wn}
  R.~Emparan and H.~S.~Reall,
  Phys.\ Rev.\ Lett.\  {\bf 88}, 101101 (2002)
  [arXiv:hep-th/0110260].

\bibitem{Mishima:2005id}
  T.~Mishima and H.~Iguchi,
  Phys.\ Rev.\  D {\bf 73}, 044030 (2006)
  [arXiv:hep-th/0504018].

\bibitem{Figueras:2005zp}
  P.~Figueras,
  JHEP {\bf 0507}, 039 (2005)
  [arXiv:hep-th/0505244].

\bibitem{Pomeransky:2006bd}
  A.~A.~Pomeransky and R.~A.~Sen'kov,
  arXiv:hep-th/0612005.

\bibitem{Elvang:2007rd}
  H.~Elvang and P.~Figueras,
  JHEP {\bf 0705}, 050 (2007)
  [arXiv:hep-th/0701035].

\bibitem{Iguchi:2007is}
  H.~Iguchi and T.~Mishima,
  Phys.\ Rev.\  D {\bf 75}, 064018 (2007)
  [arXiv:hep-th/0701043].

\bibitem{Izumi:2007qx}
  K.~Izumi,
  Prog.\ Theor.\ Phys.\  {\bf 119}, 757 (2008)
  [arXiv:0712.0902 [hep-th]].

\bibitem{Elvang:2007hs}
  H.~Elvang and M.~J.~Rodriguez,
  JHEP {\bf 0804}, 045 (2008)
  [arXiv:0712.2425 [hep-th]].

\bibitem{Evslin:2008py}
  J.~Evslin and C.~Krishnan,
  JHEP {\bf 0809}, 003 (2008)
  [arXiv:0804.4575 [hep-th]].

\bibitem{Ishibashi:2003ap}
  A.~Ishibashi and H.~Kodama,
  Prog.\ Theor.\ Phys.\  {\bf 110}, 901 (2003)
  [arXiv:hep-th/0305185].

\bibitem{Kodama:2003kk}
  H.~Kodama and A.~Ishibashi,
  Prog.\ Theor.\ Phys.\  {\bf 111}, 29 (2004)
  [arXiv:hep-th/0308128].

\bibitem{Kunduri:2006qa}
  H.~K.~Kunduri, J.~Lucietti and H.~S.~Reall,
  Phys.\ Rev.\  D {\bf 74}, 084021 (2006)
  [arXiv:hep-th/0606076].

\bibitem{Murata:2007gv}
  K.~Murata and J.~Soda,
  Class.\ Quant.\ Grav.\  {\bf 25}, 035006 (2008)
  [arXiv:0710.0221 [hep-th]].

\bibitem{Murata:2008yx}
  K.~Murata and J.~Soda,
  Prog.\ Theor.\ Phys.\  {\bf 120}, 561 (2008)
  [arXiv:0803.1371 [hep-th]].

\bibitem{Dias:2010eu}
  O.~J.~C.~Dias, P.~Figueras, R.~Monteiro, H.~S.~Reall and J.~E.~Santos,
  arXiv:1001.4527 [hep-th].

\bibitem{Kodama:2009rq}
  H.~Kodama, R.~A.~Konoplya and A.~Zhidenko,
  Phys.\ Rev.\  D {\bf 79}, 044003 (2009)
  [arXiv:0812.0445 [hep-th]].

\bibitem{Gregory:1993vy}
  R.~Gregory and R.~Laflamme,
  Phys.\ Rev.\ Lett.\  {\bf 70}, 2837 (1993)
  [arXiv:hep-th/9301052].

\bibitem{Emparan:2003sy}
  R.~Emparan and R.~C.~Myers,
  JHEP {\bf 0309}, 025 (2003)
  [arXiv:hep-th/0308056].

\bibitem{Astefanesei:2010bm}
  D.~Astefanesei, M.~J.~Rodriguez and S.~Theisen,
  arXiv:1003.2421 [hep-th].

\bibitem{Dobiasch:1981vh}
  P.~Dobiasch and D.~Maison,
  Gen.\ Rel.\ Grav.\  {\bf 14}, 231 (1982).

\bibitem{Gibbons:1985ac}
  G.~W.~Gibbons and D.~L.~Wiltshire,
  Annals Phys.\  {\bf 167}, 201 (1986)
  [Erratum-ibid.\  {\bf 176}, 393 (1987)].

\bibitem{Rasheed:1995zv}
  D.~Rasheed,
  Nucl.\ Phys.\  B {\bf 454}, 379 (1995)
  [arXiv:hep-th/9505038].

\bibitem{Larsen:1999pp}
  F.~Larsen,
  Nucl.\ Phys.\  B {\bf 575}, 211 (2000)
  [arXiv:hep-th/9909102].

\bibitem{Ishihara:2005dp}
  H.~Ishihara and K.~Matsuno,
  Prog.\ Theor.\ Phys.\  {\bf 116}, 417 (2006)
  [arXiv:hep-th/0510094].

\bibitem{Ishihara:2006iv}
  H.~Ishihara, M.~Kimura, K.~Matsuno and S.~Tomizawa,
  Class.\ Quant.\ Grav.\  {\bf 23}, 6919 (2006)
  [arXiv:hep-th/0605030].

\bibitem{Wang:2006nw}
  T.~Wang,
  Nucl.\ Phys.\  B {\bf 756}, 86 (2006)
  [arXiv:hep-th/0605048].

\bibitem{Yazadjiev:2006iv}
  S.~S.~Yazadjiev,
  Phys.\ Rev.\  D {\bf 74}, 024022 (2006)
  [arXiv:hep-th/0605271].

\bibitem{Brihaye:2006ws}
  Y.~Brihaye and E.~Radu,
  Phys.\ Lett.\  B {\bf 641}, 212 (2006)
  [arXiv:hep-th/0606228].

\bibitem{Ida:2007vi}
  D.~Ida, H.~Ishihara, M.~Kimura, K.~Matsuno, Y.~Morisawa and S.~Tomizawa,
  Class.\ Quant.\ Grav.\  {\bf 24}, 3141 (2007)
  [arXiv:hep-th/0702148].

\bibitem{Ishihara:2006ig}
  H.~Ishihara, M.~Kimura and S.~Tomizawa,
  Class.\ Quant.\ Grav.\  {\bf 23}, L89 (2006)
  [arXiv:hep-th/0609165].

\bibitem{Yoo:2007mq}
  C.~M.~Yoo, H.~Ishihara, M.~Kimura, K.~Matsuno and S.~Tomizawa,
  Class.\ Quant.\ Grav.\  {\bf 25}, 095017 (2008)
  [arXiv:0708.0708 [gr-qc]].

\bibitem{Kimura:2009gy}
  M.~Kimura, H.~Ishihara, S.~Tomizawa and C.~M.~Yoo,
  Phys.\ Rev.\  D {\bf 80}, 064030 (2009)
  [arXiv:0906.4681 [gr-qc]].

\bibitem{Matsuno:2007ts}
  K.~Matsuno, H.~Ishihara, M.~Kimura and S.~Tomizawa,
  Phys.\ Rev.\  D {\bf 76}, 104037 (2007)
  [arXiv:0707.1757 [hep-th]].

\bibitem{Nakagawa:2008rm}
  T.~Nakagawa, H.~Ishihara, K.~Matsuno and S.~Tomizawa,
  Phys.\ Rev.\  D {\bf 77}, 044040 (2008)
  [arXiv:0801.0164 [hep-th]].

\bibitem{Tomizawa:2008hw}
  S.~Tomizawa, H.~Ishihara, K.~Matsuno and T.~Nakagawa,
  Prog.\ Theor.\ Phys.\  {\bf 121}, 823 (2009)
  [arXiv:0803.3873 [hep-th]].

\bibitem{Tomizawa:2008rh}
  S.~Tomizawa and A.~Ishibashi,
  Class.\ Quant.\ Grav.\  {\bf 25}, 245007 (2008)
  [arXiv:0807.1564 [hep-th]].

\bibitem{Stelea:2008tt}
  C.~Stelea, K.~Schleich and D.~Witt,
  Phys.\ Rev.\  D {\bf 78}, 124006 (2008)
  [arXiv:0807.4338 [hep-th]].

\bibitem{Tomizawa:2008qr}
  S.~Tomizawa, Y.~Yasui and Y.~Morisawa,
  Class.\ Quant.\ Grav.\  {\bf 26}, 145006 (2009)
  [arXiv:0809.2001 [hep-th]].

\bibitem{Gal'tsov:2008sh}
  D.~V.~Gal'tsov and N.~G.~Scherbluk,
  Phys.\ Rev.\  D {\bf 79}, 064020 (2009)
  [arXiv:0812.2336 [hep-th]].

\bibitem{Allahverdizadeh:2009ay}
  M.~Allahverdizadeh and K.~Matsuno,
  Phys.\ Rev.\  D {\bf 81}, 044001 (2010)
  [arXiv:0908.2484 [hep-th]].

\bibitem{Cai:2006td}
  R.~G.~Cai, L.~M.~Cao and N.~Ohta,
  Phys.\ Lett.\  B {\bf 639}, 354 (2006)
  [arXiv:hep-th/0603197].

\bibitem{Kurita:2007hu}
  Y.~Kurita and H.~Ishihara,
  Class.\ Quant.\ Grav.\  {\bf 24}, 4525 (2007)
  [arXiv:0705.0307 [hep-th]].

\bibitem{Kurita:2008mj}
  Y.~Kurita and H.~Ishihara,
  Class.\ Quant.\ Grav.\  {\bf 25}, 085006 (2008)
  [arXiv:0801.2842 [hep-th]].

\bibitem{Ishihara:2007ni}
  H.~Ishihara and J.~Soda,
  Phys.\ Rev.\  D {\bf 76}, 064022 (2007)
  [arXiv:hep-th/0702180].

\bibitem{Chen:2007pu}
  S.~Chen, B.~Wang and R.~K.~Su,
  Phys.\ Rev.\  D {\bf 77}, 024039 (2008)
  [arXiv:0710.3240 [hep-th]].

\bibitem{Wei:2009kg}
  S.~W.~Wei, R.~Li, Y.~X.~Liu and J.~R.~Ren,
  Eur.\ Phys.\ J.\  C {\bf 65}, 281 (2010)
  [arXiv:0901.2614 [hep-th]].

\bibitem{Bizon:2006ue}
  P.~Bizon, T.~Chmaj and G.~Gibbons,
  Phys.\ Rev.\ Lett.\  {\bf 96}, 231103 (2006)
  [arXiv:gr-qc/0604043].

\bibitem{Radu:2007te}
  E.~Radu and M.~Visinescu,
  Mod.\ Phys.\ Lett.\  A {\bf 22}, 1621 (2007)
  [arXiv:0706.0992 [gr-qc]].

\bibitem{Kimura:2007cr}
  M.~Kimura, K.~Murata, H.~Ishihara and J.~Soda,
  Phys.\ Rev.\  D {\bf 77}, 064015 (2008)
  [arXiv:0712.4202 [hep-th]].

\bibitem{Ishihara:2008re}
  H.~Ishihara, M.~Kimura, R.~A.~Konoplya, K.~Murata, J.~Soda and A.~Zhidenko,
  Phys.\ Rev.\  D {\bf 77}, 084019 (2008)
  [arXiv:0802.0655 [hep-th]].

\bibitem{He:2008im}
  X.~He, B.~Wang, S.~Chen, R.~G.~Cai and C.~Y.~Lin,
  Phys.\ Lett.\  B {\bf 665}, 392 (2008)
  [arXiv:0802.2449 [hep-th]].

\bibitem{Matsuno:2009nz}
  K.~Matsuno and H.~Ishihara,
  Phys.\ Rev.\  D {\bf 80}, 104037 (2009)
  [arXiv:0909.0134 [hep-th]].

\bibitem{Peng:2009wx}
  J.~J.~Peng and S.~Q.~Wu,
  Nucl.\ Phys.\  B {\bf 828}, 273 (2010)
  [arXiv:0911.5070 [hep-th]].

\bibitem{Liu:2010wh}
  Y.~Liu, S.~Chen and J.~Jing,
  Phys.\ Rev.\  D {\bf 81}, 124017 (2010)
  [arXiv:1003.1429 [gr-qc]].

\bibitem{Hu:1974hh}
  B.~L.~Hu,
  J.\ Math.\ Phys.\  {\bf 15}, 1748 (1974).


\end{thebibliography}
\end{document}